\documentclass[11pt,a4paper]{article}
\pdfoutput=1
\usepackage{myjheppub}
\usepackage{latexsym,amsfonts,amsmath,amssymb}
\usepackage{amsthm}
\usepackage{mathtools}
\usepackage{bbm}
\usepackage{graphicx}
\usepackage{color}
\usepackage{caption}
\usepackage{subcaption}
\usepackage[normalem]{ulem}
\usepackage{comment}
\usepackage{url}
\usepackage{slashed}
\usepackage{bm}
\usepackage{hhline}

\usepackage{tabu}

\usepackage{tikz}

\newtheorem{thm}{Theorem}[section]

\newtheorem{proposition}{Proposition}[section]

\newcommand{\CN}{\mathcal{N}}
\newcommand{\CO}{\mathcal{O}}

\renewcommand{\Re}{{\rm Re}}
\newcommand{\Tr}{\mbox{Tr}}

\newcommand{\IC}{\mathbb{C}}
\newcommand{\IZ}{\mathbb{Z}}

\newcommand{\IN}{\mathbb{N}}

\newcommand{\rme}{{\rm e}}
\newcommand{\ii}{\mathrm{i}}

\newcommand\be{\begin{equation}}
\newcommand\ee{\end{equation}}
\newcommand\bea{\begin{eqnarray}}
\newcommand\eea{\end{eqnarray}}

\newcommand{\dd}{\mathrm{d}}

\renewcommand{\=}{\;= \;}
\newcommand{\+}{\;+ \;}

\renewcommand{\a}{\alpha}
\renewcommand{\b}{\beta}

\newcommand{\g}{\gamma}
\newcommand\G{\Gamma}

\newcommand{\wh}{\widehat}
\newcommand{\wt}{\widetilde}

\renewcommand{\Re}{\text{Re}}

\newcommand{\p}{\partial}

\newcommand{\ndt}{\noindent}

\newcommand{\half}{\tfrac12}
\newcommand{\z}{\zeta}

\newcommand{\bl}{\pmb \lambda}
\newcommand{\bmu}{\pmb \mu}

\newcommand{\defeq}{\; \coloneqq \;} 

\newcommand{\II}{I}

\newcommand{\tO}{\text{O}}

\renewcommand{\Re}{\mbox{Re}}

\newcommand{\wtZ}{\wt Z}
\newcommand{\tbar}{\overline t}

\newcommand{\psibar}{\overline \psi}
\newcommand{\Psibar}{\overline \Psi}

\newcommand{\vev}[1]{\left\langle{\,#1\,}\right\rangle}
\newcommand{\ket}[1]{\mid \! #1 \, \rangle}
\newcommand{\bra}[1]{\langle \, #1 \! \mid}

\newcommand{\braket}[2]{\langle \, #1 \! \mid \! #2 \, \rangle}

\newcommand{\GG}{G}

\newcommand{\tikhole}[1][0.15]{\tikz[baseline=-0.5ex]\draw[thick][black,radius=#1] (0,0) circle ;}%
\newcommand{\tikelec}[1][0.15]{\tikz[baseline=-0.5ex]\draw[black,fill=black,radius=#1] (0,0) circle ;}%

\title{Unitary matrix models, free fermion ensembles, and \\ the giant graviton expansion}

\author{Sameer Murthy}
\affiliation{Department of Mathematics, King's College London,\\
The Strand, London WC2R 2LS, U.K.}

\emailAdd{sameer.murthy@kcl.ac.uk}

\abstract{
We consider a class of matrix integrals over the unitary group~$U(N)$ with an infinite set of 
couplings  characterized by a series~$f(q)=\sum_{n \ge 1} a_n \, q^n$, with~\hbox{$a_n \in \IZ$}. 
Such integrals arise in physics as the partition functions of free four-dimensional gauge theories 
on~$S^3$ and, in particular, as the superconformal index of super Yang-Mills theory.
We show that any such model can be expressed in terms of a system of free fermions
in an ensemble parameterized by the infinite set of couplings. Integrating out the fermions 
in a given quantum state  leads to a convergent expansion as a series of determinants, 
as shown by Borodin-Okounkov many years ago. 
By further averaging over the ensemble, we obtain a formula for the matrix integral as 
a~$q$-series with successive terms suppressed by~$q^{\a N + \b}$ where~$\a$,~$\b$ 
do not depend on~$N$. This provides a matrix-model explanation of the giant graviton 
expansion that has been observed recently in the literature. 
}

\dedicated{This article is dedicated to Don Zagier on the occasion of his $70^\text{th}$ birthday.}

\begin{document}

\maketitle 

\emph{The influence of Don Zagier's work on the topics discussed in this article---matrix 
integrals, $q$-series, sum over partitions---has been deep and wide-ranging. 
Don's love and enthusiasm for interesting patterns from any area of mathematics 
is deeply infectious, and his ability to connect different parts of mathematics 
(and physics!) is a rich inspiration for many of us. This influence and inspiration 
underlie a great deal of the following text, and it is the author's hope that they are, 
at least in part, recognizable to the reader.}

\section{Introduction, philosophy, and the main statement}

The idea that matrix models---integrals over the space of matrices of a certain kind 
(hermitian/orthogonal/unitary/$\dots$)---are related to systems of fermions is an 
old one (see e.g.~\cite{Klebanov:1991qa, Ginsparg:1993is, DiFrancesco:1993cyw} 
for early reviews), and has taken multiple avatars over the years. 
Each incarnation has brought a slightly new point of view, but the basic idea can be 
understood by the fact that the appropriate measure on the space of matrices is 
proportional to the Vandermonde polynomial of the eigenvalues of the matrix, which 
vanishes whenever two eigenvalues coincide. Conflating this fact with the Pauli 
exclusion principle---that fermionic wavefunctions vanish when two fermions 
collide in phase space---leads to the identification of the matrix eigenvalues with the 
fermions (or bilinears of the fermions). 
In this article we discuss another relation between a class of unitary matrix integrals and 
a free fermionic theory coupled linearly to an infinite set of bosonic quantum variables.

Consider the following integral over the space of~$U(N)$ matrices with the invariant 
measure~$\dd U$, normalized such that the volume of the whole space is~1, 
\be \label{Uact}
Z_N({\bf g}) \= 
\int_{U(N)} \, \dd U\, \exp \biggl( \; \sum_{k=1}^\infty \, \frac{1}{k} \, g_k \,
 \Tr \, U^k \, \Tr \, U^{-k} \, \biggr) \,,
\ee
where~${\bf g} = (g_1, g_2, \dots)$ is an infinite-dimensional vector of variables, 
or \emph{coupling constants}, whose choice defines the model. 
Such integrals arise in the algebraic problem of counting
invariants of multiple matrices under simultaneous conjugation 
by~$U(N)$~\cite{Razmyslov,Procesi,Teranishi}.\footnote{Replacing the adjoint 
character of~$U(N)$ in~\eqref{Uact} by other characters also leads to a class of 
interesting problems, we will not discuss them here.} In physics, we encounter 
the very same problem as counting gauge-invariant operators in free Yang-Mills 
theory coupled to matter fields~\cite{Sundborg:1999ue, Polyakov:2001af, Aharony:2003sx}.
In this setting the coupling constants are given in terms of a power series as
\be \label{defis}
g_k\= f(q^k) \,, \quad k = 1,2,\dots \,, \qquad \text{with}
\quad f (q) \= \sum_{n=1}^\infty \,a_n \, q^n \; \in \; \IZ[[q]] \, .
\ee
(Here, and below, we use the usual notation~$\IZ[[q]]$ for the ring of power series 
in~$q$ with integer coefficients.) The resulting integral then also admits a power 
series expansion with integer coefficients 
\be  \label{indtrace}
\II^f_N (q) \= Z_N \bigl((f(q), f(q^2), \dots )\bigr) 
 \= \sum_{\ell \ge 0} d^f_N(\ell) \, q^{\ell} \= 1 + \tO(q) \,,
\ee
where the integers~$d_N(\ell)$ are given by a trace of some operator over the Hilbert space 
with a fixed charge~$\ell$. In the simplest case it is simply the dimension of that Hilbert space 
or the number of invariant polynomials of a given degree~$\ell$ in the algebraic problem. 
(The constant term in~\eqref{indtrace} is calculated by setting all the coupling constants~$g_k=0$ 
in~\eqref{Uact}.)

A particularly interesting class of generating functions of the sort~\eqref{indtrace} is 
given by \emph{superconformal indices} in four dimensional supersymmetric gauge 
theory\footnote{In fact, the most general superconformal index depends on 
multiple~$q$-type variables, but here we consider the simplest situation with one~$q$ 
variable.}, which is the generating function of the Witten index of supersymmetric (BPS) 
states that preserves some fraction of the supersymmetries. 
For each type of BPS state,~$f(q)$---called the \emph{single-letter index} in this 
context---is calculated simply only from the knowledge of the field content of the gauge 
theory~\cite{Romelsberger:2005eg, Kinney:2005ej}. 
In order to have concrete examples, we specify the theory to be~$\CN=4$ super 
Yang-Mills theory,  which has 16 complex supercharges. 
The single-letter index~$f(q)$ then takes the following values for different types of BPS 
states\footnote{To make the notation slightly lighter, we will 
use subscripts or superscripts~$i/16$ on the various functions 
corresponding to the~$\frac{i}{16}$-BPS indices. 
For example we have~$I^{i/16}_N(q)$ to mean~$I^{f_{i/16}}_N(q)$. One can also study 
the~$\frac14$-BPS index, we will not do so in this article.} 
\be \label{1overnBPS}
\begin{split}
\text{$\frac12$-BPS}:~ \quad f(q) \; \mapsto \; f_{1/2}(q) & \=   q \,, \\
\text{$\frac18$-BPS}:~ \quad f(q) \; \mapsto \; f_{1/8}(q) & \=  \frac{2q}{1+q} 
\= 2q - 2q^2 + 2q^3 - 2q^4 + 2q^5 - 2q^6 + \cdots \,, \\
\text{$\frac{1}{16}$-BPS}: \quad f(q) \; \mapsto \;  f_{1/16}(q) & \=   1-\frac{(1-q^2)^3}{(1-q^3)^2} 
\=  3q^2 - 2q^3 - 3q^4 + 6q^5 - 2q^6 \+ \cdots \,. 
\end{split}
\ee

The above physics discussion about the superconformal index can be explained better 
and made more rigorous, but we will not do so here partly because there are nice 
mathematical expositions available  (see e.g.~\cite{Spiridonov:2009za}), and partly 
because it can be taken to be a black box which generates interesting examples 
illustrating the general presentation.
There is, however, one important piece of physics that cannot really be made more rigorous
but is central to the story, namely the profound AdS/CFT conjecture (which is really an 
infinite-variable generating function of conjectures). 
One point of philosophy that Don Zagier has taught us repeatedly through his works and 
expositions  is that we should test good conjectures as extensively as we possibly can to 
make sure that it contains Truth, and that the best conjectures are beautiful and surprising 
and perhaps even contain an element of the  
outrageous. 

As a whole community of string theorists can attest to,  AdS/CFT, or more generally the idea 
of holography, has fulfilled the above criteria time and again. In situations where one has been 
able to make its predictions well-defined---even if not rigorous---from a mathematical point of 
view, it has created very beautiful and surprising mathematical structures. 
A very partial list of examples 
is: large-$N$ expansions in matrix models~\cite{tHooft:1973alw}, 
Chern-Simons theory and~CFT$_2$ (with some hindsight)~\cite{Witten:1988hf}, 
low-dimensional string theory~\cite{Witten:1990hr, Kontsevich:1992ti}
and Gopakumar-Vafa theory~\cite{Gopakumar:1998ki}. 
It is the author's belief that there are many more mathematical gems hiding within the 
structure of AdS/CFT, which are yet to be found. The aim of this article is to set up 
a dig in one particular corner. 

\vskip 0.4cm

\ndt {\bf What are we looking for?} 
\vskip 0.1cm

As we mentioned above, the integer coefficients~$d^f_N(\ell)$ in~\eqref{indtrace} 
count the index of BPS states in a four-dimensional gauge theory
of charge~$\ell$. They can be calculated to any required order on the computer 
by calculating the integral~\eqref{Uact}. (There is a more efficient method of calculation 
by writing the integral as a sum over partitions that we mention in the following.) 
From the point of view of the AdS/CFT conjecture, the integers~$d^f_N(\ell)$ 
also count the same index of BPS states of charge~$\ell$ in the dual theory of gravity 
in five dimensional asymptotically AdS space with Newton's constant~$G=1/N^2$.

Now, on the one hand, our understanding of the nature of quantum gravitational 
BPS states (and our ability to count them) is quite backward compared to say gauge 
theory, as it requires a precise definition of a quantum theory of gravity about which 
we know very little. On the other hand, we do have a good knowledge of classical 
non-linear solutions of the gravitational theory\footnote{This involves solving 
Einstein's equations with particular sources specified in the gravitational theory.}, 
like gravitons, D-branes, and black holes, which 
give rise to interesting predictions for patterns in the numbers~$d^f_N(\ell)$ in different 
regimes of the charge~$\ell$. As~$N \to \infty$, gravitons are associated with states 
of charge~$\tO(1)$, D-branes are associated with states of charge~$\tO(N)$, and 
black holes are associated with states of charge~$\tO(N^2)$. 
The simplest predictions arise by estimating the thermodynamic entropy of the 
gravitational objects, which we now briefly review in turn.

\vskip 0.2cm

\ndt \emph{Gravitons } 
Since gravitons have small energies in units of Newton's constant~$1/N^2$, we 
can quantize them using usual methods of quantum field theory, and calculate 
their statistical entropy. One can therefore also calculate the index of multi-graviton 
states in AdS$_5$ space, whose generating function is~\cite{Aharony:2003sx} 
\be \label{Imultigrav}
I^f_\text{multi-graviton}(q) \= \prod_{k=1}^\infty \frac{1}{1-f(q^k)} \,,
\ee
for all the BPS single-letter indices in~\eqref{1overnBPS}. 
The AdS/CFT conjecture says that the number~$d^f_N(\ell)$ for fixed~$\ell$ 
should have a limit~$N \to \infty$, and that the limiting value 
should be the index of multi-gravitons in AdS$_5$ space which is 
given by the~$q^\ell$ coefficient of~$I^f_\text{multi-graviton}(q)$. 
This can be proved for arbitrary~$f(q)$ by a simple argument, which we review in 
Section~\ref{sec:partitionsum}. 

\vskip 0.2cm

\ndt \emph{Black holes } 
Since black holes have energies comparable to~$N^2$ we cannot quantize them 
by any known method, and so we cannot count their statistical entropy. However, 
the profound arguments of Bekenstein and Hawking in the 1970s showed that black 
holes carry thermodynamic entropy proportional to their horizon area and that, upon 
combining this fact with the Boltzmann equation, there should be a statistical entropy 
associated to the black hole. 
In our context, the~$\frac1{16}$-BPS case is the only one where smooth black holes of 
finite horizon area are known~\cite{Gutowski:2004yv}, and the Bekenstein-Hawking 
entropy calculation leads to the prediction\footnote{This equation will be more familiar 
to physicists after recalling that Newton's constant~$G_N=1/N^2$, so that, in terms 
of~$\nu = \ell/N^2$ (the charge in gravitational units), 
we have~$\log d^{1/16}_N = N^2 \cdot \pi \nu^{2/3} /2 \cdot 3^{1/6}$, which shows 
the expected scaling of~$N^2$ from the area law~$S_\text{BH}=$Area$/4G_N$.}
\be \label{sSBH}
\log d^{1/16}_N(\ell) \= \frac{\pi N^{2/3}}{2 \cdot 3^{1/6}} \; \ell^{2/3} \; +\; \tO(\ell^{1/3}) \,,  
\quad \ell \to \infty\,.
\ee
This prediction has recently been computationally verified to large orders and then 
proved in the last few years~\cite{Cabo-Bizet:2018ehj, Choi:2018hmj, Benini:2018ywd,Cabo-Bizet:2019osg,Murthy:2020rbd},
and in this process some unexpected relations to number-theoretic functions have 
appeared~\cite{Cabo-Bizet:2019eaf,Cabo-Bizet:2020nkr}. 
We will not discuss black holes in more detail in this article.

\vskip 0.2cm

\ndt \emph{D-branes and the main statement of this article } 
The new point of the present article concerns the intermediate range of energies scaling 
linearly with~$N$, corresponding to D-branes, which are also called \emph{giant gravitons} 
in AdS space~\cite{McGreevy:2000cw,Hashimoto:2000zp,Grisaru:2000zn} 
in our context. 
Recall that the~\emph{order} of power series~$f(q)$ is the smallest positive integer~$\a$ 
such that the coefficient of~$q^\a$ is non-zero.
We show in Section~\ref{sec:giants} that for arbitrary~$f(q)$ of order~$\a>0$ 
the integral~$I^f_N$ in~\eqref{indtrace} can be written as an infinite sum of power series of 
successively higher order, 
\be \label{IGexp}
\frac{I^f_N(q)}{I^f_\infty(q)} \= \sum_{m=0}^\infty G^{(m)}_{f,N}(q) \,, \qquad 
G^{(m)}_{f,N}(q) \; \in \;  q^{\a mN +\b(m)} \, \IZ[[q]] \,.
\ee 

This type of formula was numerically observed recently in~\cite{Arai:2019xmp,Imamura:2021ytr} 
for the superconformal index, and in~\cite{Gaiotto:2021xce} for a class of 
examples with multiple charges where one of the charges scales as~$N$ and the other 
charges are fixed. The formula is then interpreted holographically in these interesting papers, 
with the~$m^\text{th}$ term in~\eqref{IGexp} identified with the index of~$m$ giant gravitons 
in the dual AdS space. In~\cite{Arai:2019xmp,Imamura:2021ytr} this identification is based 
on an explicit calculation of the index of wrapped D-branes in AdS space for the first few 
values of~$m$. In~\cite{Gaiotto:2021xce} it is based on the calculation of determinant 
operators of large powers of one matrix perturbed by other 
matrices~\cite{Berenstein:2002ke,Balasubramanian:2002sa} 
and by using the fact that giant gravitons are determinant operators in the gauge 
theory~\cite{Balasubramanian:2001nh,Corley:2001zk,Berenstein:2004kk}.

\vskip 0.2cm

\ndt {\bf Our method and results } 
In the present article we work purely with the matrix integral discussed above. 
We give an explicit formula in Section~\ref{sec:giants} for~$G^{(m)}_{f,N}(q)$ as 
a certain integral transform of a determinant of a~$m \times m$ matrix. The basic 
idea in physics language is well-known. Consider a theory of a complex massive 
fermion with a quartic interaction. This can be written as a non-interacting (or~\emph{free}) 
fermion theory coupled to a boson, using the Hubbard-Stratonovich (H-S) transformation. 
Integrating out the fermions leads to a functional determinant over the Grassman fields, and 
the original partition function is obtained by further averaging over the boson with a Gaussian 
weight. 
Here we have an infinite set of Grassmann variables, which is equivalent to a 
two-dimensional chiral fermionic field. We rewrite our integral~\eqref{Uact} as 
the average~$Z_N({\bf g})=\bigl{\langle} \wt Z_N \bigr{\rangle}_{\bf g}$ over the 
infinite set of couplings~${\bf t}$ of an auxiliary matrix model~\eqref{Uactaux}. 
From a result of~\cite{OkRandPart,BorOk}, the auxiliary model is exactly equivalent 
to a certain expectation value in a theory of free fermions in a state~$\ket{{\bf t}}$. 
The H-S transformation in this context 
is an ensemble average in the fermionic theory, controlled by the couplings~${\bf g}$.  
When~$g_k=f(q^k)$ the averaging procedure leads to an expression for~$G^{(m)}_{f,N}(q)$.  
The results are presented as Theorems~\ref{thm:giants},~\ref{thm:onegiant}.

It is convenient to translate the original matrix integral into a sum over partitions of 
integers with a given weight, as we review in Section~\ref{sec:partitionsum}. This 
step naturally discretizes the problem (without losing any information) and makes 
it much easier to study the various power series on a computer. It also allows us to 
make contact with the beautiful work of~\cite{OkRandPart,BorOk}, as we review in 
Section~\ref{sec:determinants} using the language of two-dimensional fermions and 
bosonization. 
In particular, they give an explicit formula for the fermionic determinant as a series of 
determinants with successively increasing number of Fourier modes of the fermion 
(0-point function + 2-point function + 4-point function~$+\dots$). 
In order to study our original theory of interest, we need the integral transform 
(the H-S transformation), which we review in Section~\ref{sec:HStrans}.
As we explain in Section~\ref{sec:giants}, the transform of the fermionic~$2m$-point 
function is~$\tO(q^{\a m N+\b})$, and is identified as the contribution of~$m$ giant-gravitons.
Remarkably, the observables of the giant graviton theory are described in terms of 
the following kernel for 2-point functions 
\be
\wh K_f(\z,q) \= \frac{1}{(1-\z)(1-1/\z)} \prod_{n=1}^\infty 
\biggl(\frac{(1-q^n)^{2} }{(1-q^n \z) (1-q^n /\z)} \biggr)^{\wh a_n} \,,
\ee
where~$\wh a_n$ are the coefficients of the \emph{dual single-letter index}~$\wh f(q)$ given 
by\footnote{The expansions given in~\cite{Gaiotto:2021xce} also involve a dual index~$\wt f$ 
related to the original~$f$ through an analytic continuation of the chemical potential. 
The precise relation of~$\wt f$ to our~$\wh f$ is not immediately clear to us.}
\be \label{isdualseries}
\wh f(q) \= \frac{f(q)}{1-f(q)} \,, \qquad \wh f(q) \= \sum_{n=1}^\infty \, \wh a_n \, q^n \,.
\ee 

\vskip 0.2cm

We end this introduction with a few comments. Firstly, we encounter infinite sums of power series
as in~\eqref{IGexp} in this article. Because the~$m^\text{th}$ term starts at~$\tO(q^{\a m N + \b})$,
the coefficient of any given power of~$q$ receives contributions from only a finite number of terms 
and is therefore well-defined. We will sometimes call such a sum over~$m$ a convergent sum in the  
power-series sense. 
Secondly, although we focus on matrix integrals that are governed by a power series~$f(q)$,
many of the assertions hold more generally in terms of the variables~${\bf g}=(g_1,g_2,\dots)$. 
In order to discuss infinite sums in this setting, it is useful to have the notion of a degree, defined 
by assigning degree~$k$ to~$g_k$ and then extending to monomials by demanding that the 
degree of the product of two variables is the sum of the individual degrees. Any power series 
in~${\bf g}$ then has a minimal degree which, on specializing~$g_k=f(q^k)$, translates to the 
order of the resulting power series in~$q$.
Now we see that an infinite sum of power series in~${\bf g}$ is well-defined if the coefficient of 
a given monomial is zero in almost all the terms of the infinite sum. This is the case if for 
any~$\rho \in \IN$, $\exists \; m=m_\rho$ such that the terms~$m, m+1, \dots$ in the sum 
are all of degree greater than~$\rho$.

Finally, we should mention that various subsets of the ideas mentioned above have been 
studied in the string theory literature,
The papers~\cite{Liu:2004vy,Alvarez-Gaume:2005dvb,AnninosSilva,Copetti:2020dil} use 
the Hubbard-Stratonovich transformation to study matrix models with a finite set of couplings. 
The relation of unitary matrices to random partitions and free fermions have been studied 
in~\cite{Dolan:2007rq,Dutta:2007ws,Ramgoolam:2016ciq,Kimura:2020cbs}, 
also mainly for reduced models. 
This approach of integrating out fermions and integrating in bosons to treat determinants
and D-branes/giant gravitons have been used fruitfully 
in~\cite{Maldacena:2004sn,Jiang:2019xdz,Gaiotto:2021xce}, 
and is summarized by the general paradigm of open-closed-open duality of~\cite{GopOCO}. 
All these developments have served as an inspiration for the present work.

\section{The partition sum as a sum over partitions \label{sec:partitionsum}}

Unitary matrix models of different types were studied in the context of two-dimensional gauge theory
in the early 
1990s~\cite{DiFrancesco:1992cn,Gross:1993hu,Minahan:1993np,Douglas:1993xv,Douglas:1993iia,Cordes:1994fc,Kazakov:1995ae}.
An important idea used in these works is to express the matrix integral as a sum over partitions of 
integers with a certain weight associated to each partition. This rewriting allows us to relate the 
theory to a string theory, and also leads to an understanding of phase transitions in the model 
upon tuning of parameters. In this section we follow this idea and express the integral~\eqref{Uact} 
as a sum over partitions, giving an independent treatment based on~\cite{Murthy:2020rbd}.  

We first set up some notations. 
We denote partitions as~$\bl = (\lambda_1, \lambda_2,\dots)$ with~$\lambda_1 \ge \lambda_2 \ge \dots $,
or in the frequency representation as~$1^{r_1} \, 2^{r_2} \, \dots$, $\lambda_j, r_i \in \IN_0$.
The number of parts of a partition (or length) is the number of non-zero~$\lambda_i$ or, 
equivalently,~$\ell(\bl) \= \sum_{i} \, r_i$. 
The weight of the partition~$|\bl| = \sum_{j\ge1} \lambda_j \= \sum_{i\ge1} \, i \, r_i$. 
To rewrite the integral~\eqref{Uact} as a sum over partitions, we first 
expand the exponential in the integrand as sums of products of traces of powers of the 
unitary matrix. We then express each product as a linear combination of~$U(N)$ characters. 
The coefficients, given by the Frobenius character formula are characters of the symmetric 
group~$S_N$, which are labelled by partitions of~$N$. Finally, we integrate over the gauge group 
using the orthogonality relation of the characters in order to obtain an expression for the 
index~$\II_N$ as a sum over representations of~$S_N$ i.e.~as a sum over partitions of~$N$. 
We review these steps in more detail in Appendix~\ref{App:sumparts}.

Upon following the steps sketched in the above paragraph, we obtain 
\be  \label{ZNpartsum}
Z_N({\bf g}) \= \sum_{\bl} \, \frac{ {\bf g}^{\bl}}{z_{\bl}} \,\sum_{\ell(\bmu)\le N} \, \chi^{\bmu}(\bl)^2 \,.
\ee
Here, and in the following, we use the following notations. 
Firstly, all sums over partitions run over all partitions (as in the sum over~$\bl$)
with any restrictions being indicated explicitly (as in the sum over~$\bmu$).
For any partition~$\bl$ as above, 
\be
{\bf g}^{\bl} \defeq  \prod_{j\ge 1} \, g_{\lambda_j} \= \prod_{i\ge 1} \, g_i^{r_i} \,, \qquad 
z_{\bl} \defeq \prod_{i} \, r_i ! \, i^{r_i} \,. 
\ee
The symbol~$\chi^{\bmu}(\bl)$ denotes the character of the symmetric group---recall that the irreps 
as well as the conjugacy classes of the symmetric group~$S_N$ are labelled by partitions of~$N$.

When~$N=\infty$, 
the sum over~$\bmu$ in~\eqref{ZNpartsum} runs over a complete set of characters, i.e.~all partitions.
In this case we can use the second orthogonality of the characters of~$S_{|\bl|}$, i.e.,
\be\label{chiorth2}
\sum_{\bmu}  \, \chi^{\bmu}(\bl) \, \chi^{\bmu}(\bl')   \=  z_{\bl}\, \delta_{\bl, \bl'} \,, 
\ee
where~$\delta_{\bl,\bmu} =1$~if~$\bl = \bmu$ and zero otherwise, 
so that 
\be \label{Iinfty}
Z_\infty({\bf g}) \= \sum_{\bl}\, {\bf g}^{\bl}  \=  \prod_{k=1}^\infty \frac{1}{1-g_k} \,.
\ee

\vskip 0.4cm

For our particular case of interest when~$g_k=f(q^k)$, we have~\cite{Murthy:2020rbd}
\be  \label{INqpartsum}
\II^f_N(q) \= \sum_{\bl} \, \frac{1}{z_{\bl}} \, f_{\bl} (q) \,\sum_{\ell(\bmu)\le N} \, \chi^{\bmu}(\bl)^2 \,,
\qquad 
f_{\bl} (q) \=  \prod_{j\ge 1} \, f(q^{\lambda_j}) \,.
\ee
Note that, because the single-letter expansion~\eqref{defis} is~$\tO(q)$, the only partitions 
that contribute to the sum over~$\bl$ in~\eqref{INqpartsum} for a fixed term~$q^{\ell_0}$ 
in the~$q$-series expansion~\eqref{indtrace} of~$\II^f_N(q)$ are those with~$|\bl| \le \ell_0$. 
Since the characters are only non-zero for~$|\bl| = |\bmu|$, the only partitions that contribute 
to the sum over~$\bmu$ obey~$|\bmu| \le \ell_0$, and since~$\ell(\bmu) \le |\bmu|$ for 
any partition, we have  that~$\ell(\bmu) \le \ell_0$.  
When~$\ell_0\le N$, the sum over~$\bmu$ can therefore be replaced by a sum over all 
partitions without changing the result, so that effectively we have~$N=\infty$, for which
the index is given by 
\be \label{Iinfty}
\II^f_\infty(q) \= \sum_{\bl}\, f_{\bl}(q)  \=  \prod_{k=1}^\infty \frac{1}{1-f(q^k)} \,,
\ee
which we write as a~$q$-series
\be \label{Iinftycoeffs}
\II^f_\infty(q) \= \sum_{\ell \ge 0} d^f_\infty(\ell) \, q^{\ell} \,.
\ee
More generally, for a power series of order~$\a>0$, 
the above arguments hold with the replacement~$\ell_0 \to \ell_0/\a$. 
(Recall that the order of the power series~$f(q)=\sum_{k\ge 0} \, a_k \, q^k$ is the power 
of the leading term, i.e.~the smallest integer~$k$ with~$a_k \neq 0$.)
We have thus proved a result 
slightly stronger than the one predicted by the AdS/CFT statement about gravitons.

\vskip 0.4cm

\begin{thm} \emph{(Graviton stability)} \\
For any single-particle index~$f(q)$ as in~\eqref{defis} of order~$\a$, the $q$-series 
coefficients in~\eqref{indtrace} of the integral defined by~\eqref{Uact},~\eqref{defis} obeys 
\be \label{dNinftyrel}
d^f_N(\ell) \=  d^f_\infty(\ell) \,, \qquad \ell \le \a N \,,
\ee 
where the~\emph{infinite-$N$} index is given by~\eqref{Iinfty},~\eqref{Iinftycoeffs}. 
\end{thm}
\vskip 0.4cm

\ndt 
\emph{Note:} The AdS/CFT prediction for graviton growth discussed around~\eqref{Imultigrav} 
is the statement of the stability---the existence of and formula for the limit---of~$d^f_N(\ell)$ 
for fixed~$\ell$ as~$N \to \infty$ for specific single-particle indices. 
That statement follows as a simple corollary of the above theorem after noting that the 
infinite-$N$ index is the same as the multi-graviton index~\eqref{Imultigrav},

\vskip 0.4cm

The statement made in the above theorem about matrix integrals should be known to mathematicians 
and physicists working in different fields. From the algebraic point of view, it is known that the ring 
of~$U(N)$ invariants is generated by all the traces of products of matrices, with possible relations 
among them due to the Cayley-Hamilton theorem~\cite{Razmyslov,Procesi,Teranishi}.
As the rank~$N \to \infty$, there are no relations and so the number of invariants at a given order 
stabilizes. 
From the physics point of view, the most interesting aspect is that an independent count of the 
number of gravitons in AdS space~\cite{Aharony:2003sx,Gunaydin:1984fk} also gives the same 
answer~\eqref{Imultigrav}. (To readers not familiar with AdS/CFT, and to see why this is non-trivial, 
we note that the graviton calculation starts with harmonic analysis on hyperbolic space
and carries a completely different flavor compared to what we are discussing here.)
This was explained in the nice paper by~\cite{Chang:2013fba} as a consequence of
the global algebras being the same on the two sides of the AdS/CFT correspondence,
and using this idea they explicitly constructed the graviton operators in the Yang-Mills theory. 

It is natural to ask if we can go beyond~\eqref{dNinftyrel} and obtain similar explicit formulas for 
the coefficients~$d^f_N(\ell)$ for~$\ell > \alpha N$ which can then be given a 
gravitational interpretation. This will be our goal, which we reach in Section~\ref{sec:giants}.
In order to reach these results, we first introduce a very useful integral transformation in the 
following section. 

\section{Hubbard-Stratonovich transformation \label{sec:HStrans}}

In this section we relate our original matrix model~\eqref{Uact} to the following auxiliary 
matrix model (which is of interest in its own right), 
\be \label{Uactaux}
\wtZ_N({\bf t}^+, {\bf t}^- ) \= 
\int_{U(N)} \, \dd U\, \exp \biggl( \; \sum_{k=1}^\infty \, \frac{1}{k} \, \Bigl( t_k^+ \,
 \Tr \, U^k \, +  t^-_k \, \Tr \, U^{-k} \,\Bigr) \biggr) \,,
\ee
where~${\bf t}^+ =(t^+_1, t^+_2, \dots)$ and~${\bf t}^- =(t^-_1, t^-_2, \dots)$ 
are two independent sets of coupling constants. This integral has also been discussed 
in different contexts by mathematicians as well as by physicists. 
In physics, this model is often 
studied together with our original model~\eqref{Uact}, one reason being that they are related 
by the \emph{Hubbard-Stratonovich transformation}~\cite{Stratonovich, Hubbard}.
Roughly speaking,
the idea is to transform a physical system with interactions (here~$\Tr \, U \, \Tr \, U^{-1}$) 
into a sum of non-interacting physical systems (here~$\Tr \, U + \Tr \, U^{-1}$). 
This is achieved by the simple trick of completing the square in a Gaussian integral. 

Let us take, for the moment,~$t, \tbar \in \IC$ to be complex conjugate variables. 
Then, for any function~$f: \IC \to \IC$ which does not grow rapidly at infinity, 
and for any~$g \in \IC$, $\Re(g) >0$, 
we introduce the averaging operator 
\be \label{defvev}
\vev{f}_g \= \int_{\IC} \frac{\dd t \, \dd \tbar}{2 \pi g} \, \rme^{-t \, \tbar/g} \, f(t) \,,
\ee
where~$\dd t \, \dd \tbar$ indicates~$\dd t \, \wedge \, \dd \tbar$ on the complex 
plane, normalized such that
\be
\vev{1}_g \= 1 \,.
\ee
In fact, the integral~\eqref{defvev} is well-defined even if~$f$ has exponential growth 
with an exponent linear in~$t$ and~$\tbar$ with proportionality constants less than~$1/g$ 
in magnitude. We can always achieve this to be the case for any given such function, 
by taking~$g$ to be small enough and then analytically continuing the result. 
This gives, for~$\a, \b \in \IC$, 
\be \label{Gaussianavg}
\vev{\exp \bigl(\a \, t  + \b \, \tbar \, \bigr)}_g \= \exp(\,g \, \a \, \b \,) \,, \qquad \a, \b \in \IC \,.
\ee
This is essentially the statement that the Fourier transform of a Gaussian is a Gaussian, 
and is used many times in the following.

Note that for~$(t^+, t^-) =(t,\tbar)$ a pair of complex conjugate variables as above, 
we have
\be \label{ttbargtrans}
\vev{(t^+)^{n_+} \; (t^-)^{\, n_- } }_g \= n! \, g^n \, \delta_{n^+, n^-} \,, \qquad n^+, n^- \in \IN_0 \,.
\ee
As is usual in quantum field theory, we use this equation to \emph{define} 
a transformation for arbitrary~$t^+, t^-$, which could be independent 
complex numbers or even formal variables.  
We then extend the operation, by linearity, to 
\be
\vev{\, \cdot \, }_g : \IZ[[{t}^+, {t}^-]] \; \longrightarrow \; \IZ[[g]] \,.
\ee
By expanding the exponential, we see that the transformation~\eqref{Gaussianavg} 
continues to hold. We continue to write the transformation in the form of a Gaussian 
integral as in~\eqref{defvev} (without specifying the range of integration) even 
when~$t^+$ and~$t^-$ are independent variables.

Finally, we extend the process inductively to any number of coupling constants labelled 
by~${\bf g} = (g_1,g_2, \dots)$. The averaging operation acts on power 
series~$f({\bf t}^+, {\bf t}^-)$ in the infinite set of variables~${\bf t}^+, {\bf t}^-$ as 
\be \label{HStrans}
\vev{ f  }_{\bf g} \defeq \prod_{k=1}^\infty \int  \frac{\dd t_k^+ \, \dd t_k^-}{2 \pi k g_k} \, 
\rme^{-t_k^+\, t_k^-/k g_k} \, f({\bf t}^+, {\bf t}^-) \,, 
\qquad \text{or} \quad \vev{\, \cdot \, }_{\bf g} \= \vev{ \vev{ \dots}_{2g_2}}_{g_1} \,.
\ee
With this set up, it is now easy to see that the integrals in~\eqref{Uact} and~\eqref{Uactaux} 
are related as 
\be \label{ZNZtlNrel} 
\vev{\wt Z_N}_{\bf g} \= Z_N({\bf g}) \,.
\ee

\vskip 0.4cm

\ndt {\bf In the language of partitions}

\vskip 0.1cm

The relation~\eqref{ZNZtlNrel} can also be understood directly from the presentation of 
the respective integrals as a sum over partitions as in Section~\ref{sec:partitionsum}.
Say we have an infinite set of variables~${\bf t} = (t_1,t_2, \dots)$. 
We define the functions  
\be \label{schurdef}
S_{\bmu}({\bf t}) \= \sum_{\bl} \, \frac{{\bf t}^{\bl}}{z_{\bl}} \, \chi^{\bmu}(\bl) \,,
\ee
where, in the notation used in Section~\ref{sec:partitionsum}, 
\be
z_{\bl} \= \prod_{i} \, r_i ! \, i^{r_i} \,, \qquad \bl \= 1^{r_1} \, 2^{r_2} \dots \,.
\ee
Note that if~${\bf t}$ is given by the power sum symmetric functions of an 
alphabet~${\bf x}=(x_1, x_2, \dots)$, i.e.~if~$t_k = \sum_{i} x_i^k$, the 
functions~\eqref{schurdef} are the Schur functions of~${\bf x}$. 
This will play a role in the following (see Appendix~\ref{App:bosonization}).
With an abuse of notation, we will therefore call the~$S_{\bmu}({\bf t})$ Schur functions.
Now, using similar manipulations as those needed to reach~\eqref{ZNpartsum1}, 
we obtain the following statement 
(which, as the author learned from~\cite{BorOk}, is part of Gessel's theorem~\cite{Gessel}), 
\be \label{ZNtlSS}
\wt Z_N({\bf t}^+, {\bf t}^-) \=  \sum_{\ell(\bmu)\le N} \, S_{\bmu}({\bf t}^+) \,  S_{\bmu}({\bf t}^-)  \,.
\ee
The~$N=\infty$ partition function is given by 
\be \label{ZNtlinfty}
\begin{split}
\wt Z_\infty({\bf t}^+, {\bf t}^-) & \=  \sum_{\bmu} \, S_{\bmu}({\bf t}^+) \,  S_{\bmu}({\bf t}^-)  
 \=  \sum_{\bmu} \sum_{\bl, \bl'} \, \frac{{{\bf t}^+}^{\bl}}{z_{\bl}} \, \frac{{{\bf t}^-}^{\bl'}}{z_{\bl'}} \, 
\chi^{\bmu}(\bl) \, \chi^{\bmu}(\bl') \\
& \=\sum_{\bl} \, \frac{1}{z_{\bl}} \, {{\bf t}^+}^{\bl}\, {{\bf t}^-}^{\bl} 
\= \exp \biggl( \, \sum_{k=1}^\infty \frac{1}{k} \, t^+_k \, t^-_k  \biggr) \,,
\end{split}
\ee
where we use the orthogonality relation~\eqref{chiorth2} in obtaining the third equality.

It is easy to see that the H-S transformation~\eqref{HStrans} acts on a product of the 
Schur functions as
\be \label{ssHStran}
\vev{S_{\bmu}({\bf t}^+) \, S_{\bmu}({\bf t}^-)}_{\bf g} \=  \sum_{\bl} \, 
\frac{{\bf g}^{\bl}}{z_{\bl}} \, \chi^{\bmu}(\bl)^2 \,.
\ee
Upon putting together the equations~\eqref{ssHStran} and~\eqref{ZNtlSS} we obtain 
\be
\vev{\wt Z_N}_{\bf g} \= \sum_{\bl} \, \frac{ {\bf g}^{\bl}}{z_{\bl}} \,\sum_{\ell(\bmu)\le N} \, \chi^{\bmu}(\bl)^2 \,, 
\ee
and recalling from Equation~\eqref{ZNpartsum} that the right-hand side of the above equation 
is precisely~$Z_N({\bf g})$, we are led once again to the relation~\eqref{ZNZtlNrel}.

\section{The determinant expansion \label{sec:determinants}}

In this section we present an explicit formula for the partition function~$\wtZ_N({\bf t}^+, {\bf t}^- )$
as a sum over Fredholm determinants.
This whole section is a review of~\cite{BorOk} where such a formula was presented based on 
the ideas of random partitions and the \emph{infinite wedge representation}~\cite{OkRandPart}  
in Vertex Operator Algebras~\cite{kac_1990}. 
We stay close to the original presentation of~\cite{BorOk}, but with some linguistic variations, 
partly because of the author's mother tongue, and partly because of his feeling that this beautiful 
work is not sufficiently known in the physics community.\footnote{Aspects of random partition 
theory have been used very productively in obtaining the powerful results on~$\CN=2$ gauge 
theory~\cite{Nekrasov:2002qd,Nekrasov:2003rj}, as well as in the ideas of quantum foam. 
See also~\cite{Betzios:2017yms}.
However, the particular matrix model we study here does not seem to have received 
too much attention in the physics literature.}$^,$\footnote{We refer the more mathematically 
minded reader to the original references~\cite{kac_1990,OkRandPart} with the agreement of 
meeting again at the end of the section.} 
We will thus use the physics language of free chiral fermions, bosonization, and their associated 
conventions, e.g.~as summarized in~\cite{Polchinski:1998rr}. Furthermore, when the matrix model 
arises from Yang-Mills theory, it is clear that the underlying chiral fermionic quantum field theory 
should really be thought of as the theory of the BPS subsector of the gauge theory. 
It would be interesting if there are precise relations with~\cite{Lin:2004nb} and with~\cite{Beem:2013sza}.

\vskip 0.4cm

\ndt {\bf A review of free fermions}

\vskip 0.1cm

First we need to set up the formalism of free fermions. 
We consider a complex chiral fermionic field~($\psi$,~$\psibar$) on the complex plane. 
(Here~$\psibar$ is the Hermitian conjugate of~$\psi$, and the chiral property in this context 
means that they are both meromorphic functions of~$z$.)
The fields have a Laurent expansion (corresponding to the NS sector), 
\be
\psi(z) \=  \sum_{r \in \IZ+\half}  \frac{\psi_r}{z^{r+1/2}} \,, \qquad 
\psibar(z)\= \sum_{r \in \IZ+\half}  \frac{\psibar_r}{z^{r+1/2}}  \,.
\ee
In the quantum theory the fields are interpreted as vertex operators obeying the algebra
\be \label{psiOPE}
 \psi(z) \, \psibar(w) \sim \frac{1}{z-w} + \tO(1)\,, \quad \psi(z) \,\psi(w) \=  \tO(z-w)  \,,
 \quad \psibar(z) \,\psibar(w) \= \tO(z-w)  
\ee
as~$z-w \to 0$, $|w| < |z|$
Equivalently, their Laurent coefficients obey the Clifford algebra, 
\be\label{psirsanticomm}
 \{\psi_r,\psibar_s \} \= \delta_{r+s,0} \,, \qquad \{\psi_r,\psi_s \} \= 0 \,, 
 \qquad \{\psibar_r,\psibar_s \} \= 0 \,, \qquad r,s \in \IZ + \half \,,
\ee
and obey the  Hermitian conjugation relations~$\psi_r^\dagger \= \psibar_{-r}$.

The Hilbert space is a representation of the algebra~\eqref{psirsanticomm} and is constructed 
as follows. For each~$r \in \IZ + \half$ (called the~\emph{energy level}), we first construct the 
two-dimensional representation of the Clifford algebra given by setting~$s=-r$ in~\eqref{psirsanticomm}. 
We denote the two basis states at every energy level as an \emph{electron} $\ket{\tikelec}_r$ or 
a \emph{hole} $\ket{\tikhole}_r$ (the lack of an electron) with the fermionic operators acting as 
\be
\ket{\tikhole}_r \; \stackrel{\psi_{-r}}{\longmapsto} \; 
\ket{\tikelec}_r \; \stackrel{\psibar_{r}}{\longmapsto} \; \ket{\tikhole}_r \,, 
\qquad r \in \IZ + \half \,.
\ee
The full Hilbert space (fermionic Fock space) is the infinite-dimensional wedge product 
of the two-dimensional representations associated to each~$k \in \IZ + \half$, constructed 
as follows. There is one special state of the system called the Dirac vacuum $\ket{0}$ 
\be
\psi_r \ket{0} \= 0 \,, \qquad \psibar_{r} \ket{0} \= 0 \,, \qquad r > 0 \,.
\ee
In the electron/hole picture, the vacuum state is populated by electrons at all negative energy 
levels and holes at all positive energy levels, because of which it is also called the 
\emph{Fermi sea} of electrons. The states of the Hilbert space are states of the form 
\be \label{excitations}
\ket{r_1,\dots r_m \,; \, \overline{r}_1,\dots \overline{r}_{\overline{m}}} 
\= \prod_{i=1}^{m} \psi_{-r_i} \, \prod_{j=1}^{\overline{m}} \psibar_{-\overline{r}_j} \, \ket{0} \,, 
\qquad r_i,  \overline{r}_j > 0\,,
\ee
for some values of~$m, \overline{m} \ge 0$. 
The inner product of two states~$\ket{a}$ and~$\ket{b}$ in the Hilbert space is denoted 
by~$\braket{a}{b}$, and the Dirac vacuum is taken to obey~$\braket{0}{0}=1$. The 
algebra~\eqref{psirsanticomm}, together with the Hermitian conjugation relation, implies that 
the basis vectors~\eqref{excitations} are orthonormal.

The states~\eqref{excitations} are finite excitations around the {Fermi sea}---one can 
remove a finite number~$\overline{m}$ of electrons (or, equivalently, create~$m$ holes) 
inside the Fermi sea, and excite a finite number~$m$ electrons at the levels above the 
Fermi sea. (In the following, we will only be concerned with states with an equal 
number~$m=\overline{m}$ of excited electrons and holes.) 
We will say that an energy level is part of a state if it is excited in that state, i.e.,
for~$r \in \IZ + \half$ and~$\ket{\psi}$ of the type~\eqref{excitations}, 
\be
r \, \in \, \ket{\psi} \quad 
\Leftrightarrow 
\quad \text{$r=r_i$ or~$r=\overline{r}_i$, for some~$i$.}
\ee
At any energy level, the occupation-number operator at level~$r \in \IZ+\half$
\be \label{defNr}
N_r \defeq 
\begin{cases} 
\psibar_{-r} \, \psi_r \,, \quad r >0 \,, \\ 
\psi_{-r} \, \psibar_r \,, \quad r < 0 \,, \\ 
\end{cases} 
\ee 
measures whether that level is excited or not compared to the Dirac vacuum, i.e.,
\be
N_r \ket{\psi} \= 
\begin{cases} 
1 \,, \quad r \, \in \, \ket{\psi} \,, \\ 
0 \,, \quad r \,\notin \, \ket{\psi}  \,. \\ 
\end{cases} 
\ee

\vskip 0.4cm

\ndt {\bf Partitions as free fermionic states}

\vskip 0.1cm

\begin{figure}\centering
\includegraphics[width=10cm]{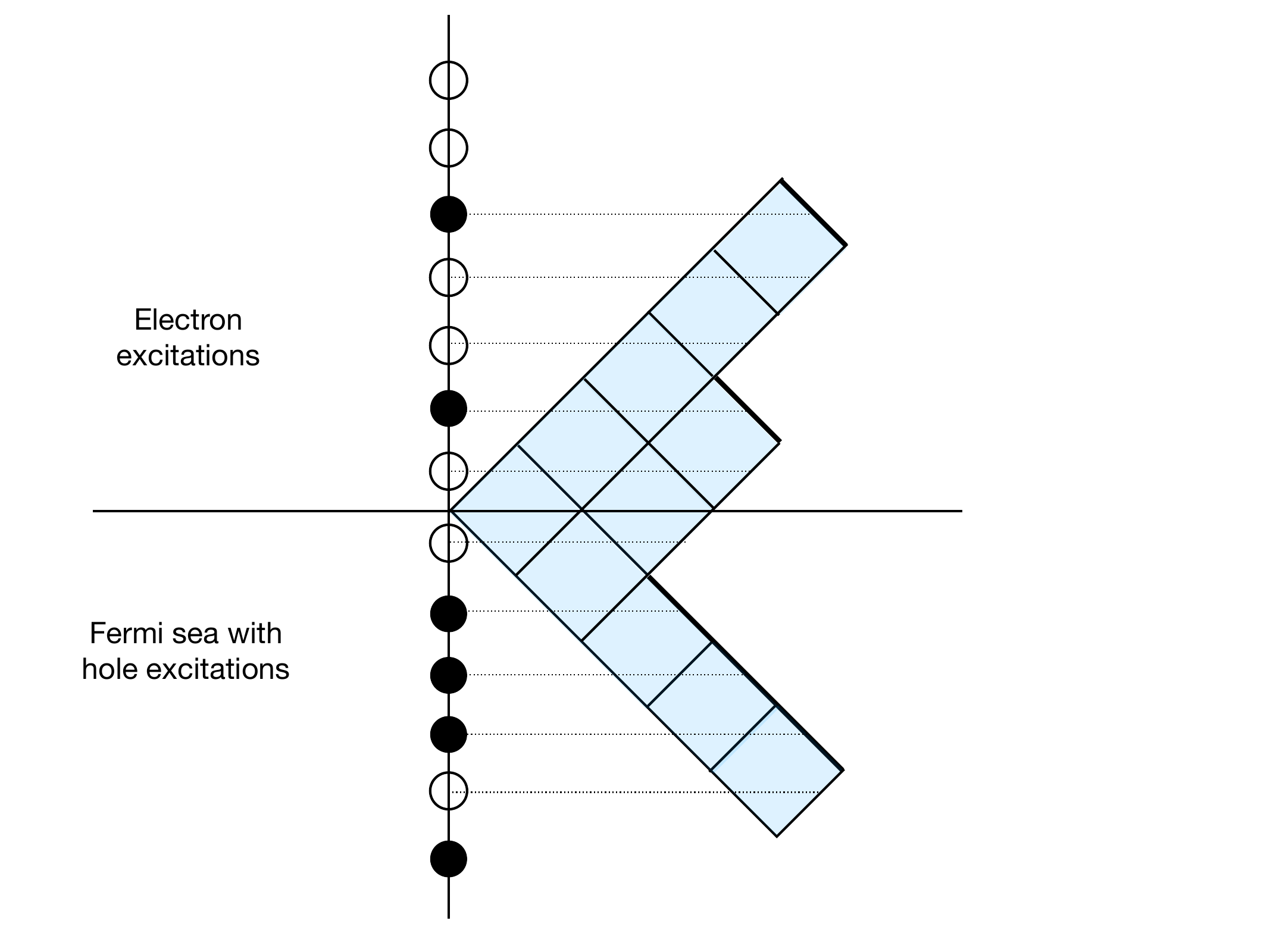}
  \caption{ Each partition is associated with a state in the free-fermion Hilbert space. 
  Here the partition~$(5,3,1,1,1)$ (rotated by $45^\circ$) 
  is associated with~$\psi_{-\frac32} \, \psi_{-\frac92} \, \psibar_{-\frac12} \, \psibar_{-\frac92} \,\ket{0} $ }
  \label{Partitions}
\end{figure}

Now, what is the relation between free fermions and our main problem? 
The starting point, as introduced in~\cite{OkRandPart}, is that there is a bijection 
between partitions and states in the Hilbert space of the free fermion. In order to introduce 
this bijection, recall that the Frobenius coordinates of any 
partition~$\bmu = \mu_1 \ge \mu_2 \ge \dots $ (modified as in~\cite{ZagBlochOkounkov} 
by adding a half) are defined as
\be
\a_i \= \mu_i - i + \half \,, \qquad \b_i \= \mu^{\text t}_i - i + \half \,, \qquad i = 1,2,\dots s \,,
\ee 
where~$s$ is the length of the largest principal diagonal of the Young diagram 
corresponding to~$\bmu$, and~${\bmu}^{\text{t}}$ is the partition conjugate to~$\bmu$. 
These coordinates obey
\be
\a_1 > \a_2 > \dots > \a_s > 0\,, \qquad \b_1 > \b_2 > \dots > \b_s > 0   \,.
\ee
We now associate to such a partition the fermionic state 
\be \label{partstate}
\ket{\bmu} \= \prod_{i=1}^s \psi_{-\a_i} \, \psibar_{-\b_i} \ket{0} \,.
\ee
It is an easy consequence of the orthonormality of~\eqref{excitations} that the 
states~\eqref{partstate} are also orthonormal.
In the Fermi sea picture, the state~$\ket{\bmu}$ corresponds to excitations at 
levels~$\a_i$ (electrons) and at~$\b_i$ (holes), $i=1,\dots ,s$, and has the intuitive 
understanding shown in Figure~\ref{Partitions}.
\footnote{We rotated this picture by~$90^\circ$ compared to~\cite{OkRandPart}
so as to represent the Fermi level horizontally.} 
In this basis,~$r \in \ket{\bmu}$ simply means that~$r$ is one of the modified 
Frobenius coordinates of~$\bmu$.

\vskip 0.4cm

\ndt {\bf The auxiliary matrix model and the determinantal expansion}

\vskip 0.1cm

We are now ready to translate our problem to the language of free fermions. 
For any~${\bf t}=(t_1,t_2,\dots)$ as in~Section~\ref{HStrans},
we define the following state in the fermionic Fock space,
\be \label{defkett}
\ket{{\bf t}} \defeq \sum_{\bmu} S_{\bmu}({\bf t}) \ket{{\bmu}} \,,
\ee
where~$S_{\bmu}$ is the Schur function defined in~\eqref{schurdef}, and  
the sum runs over all partitions as in the previous sections. 
The inner product between two such states is 
\be
\braket{{\bf t}^+}{{\bf t}^-}  \= \sum_{\bmu} S_{\bmu}({\bf t}^+) \,  
S_{\bmu}({\bf t}^-) \= \wt Z_\infty({\bf t}^+, {\bf t}^-)\,,
\ee
which is precisely the infinite-$N$ partition function~\eqref{ZNtlSS} of the auxiliary model.

\begin{figure} 
\centering
\includegraphics[width=8cm]{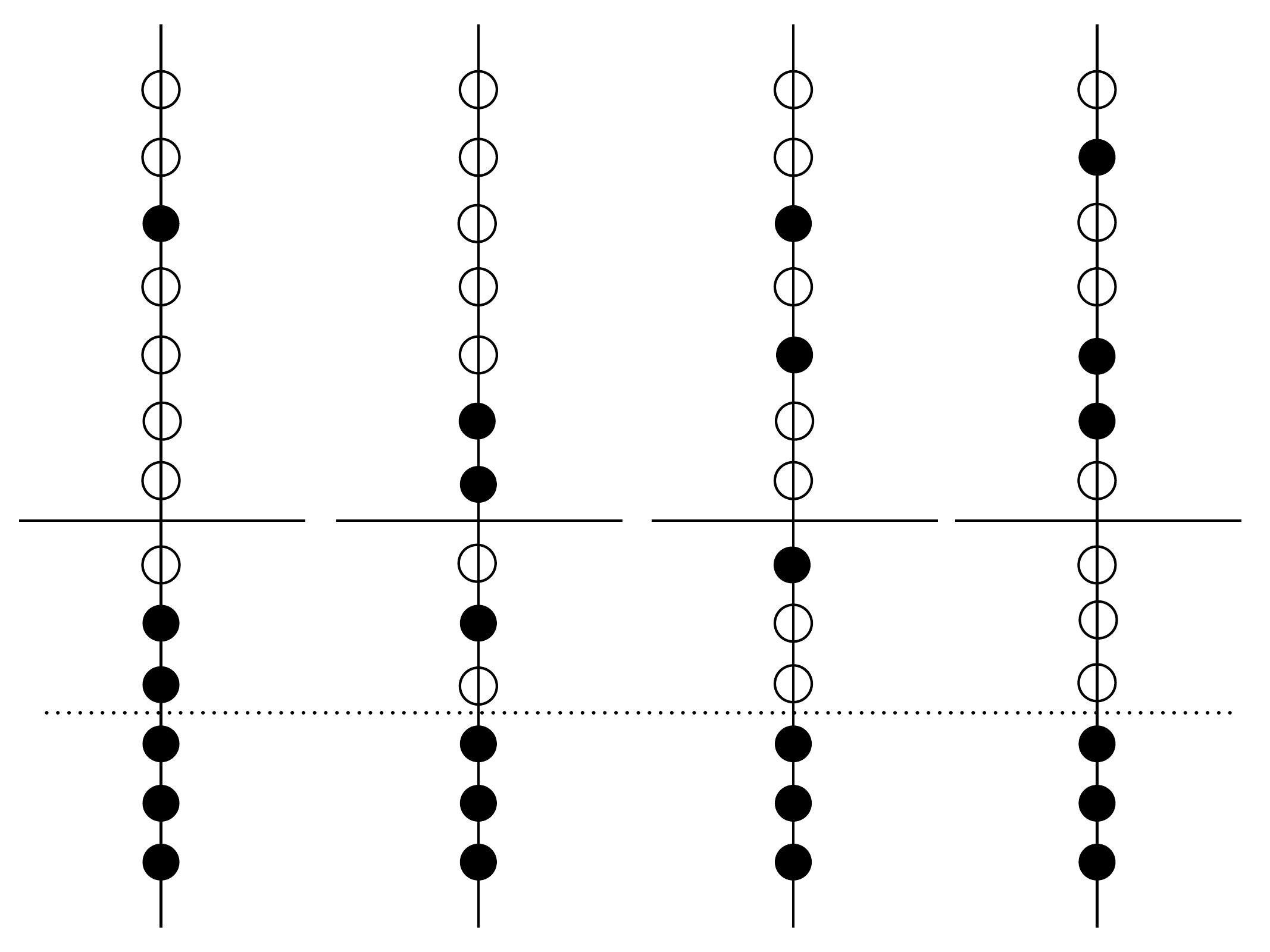}
  \caption{\emph{There is no hole in the bottom of the sea.} For finite~$N$, we can excite 
  any number of holes at depths less than~$N$ in the Fermi sea and can excite the same 
  number of electrons at arbitrary heights above the sea. Some allowed configurations are 
  shown for~$N=3$.}
  \label{FiniteN}
\end{figure}

Now we come to finite~$N$. 
Firstly, we note that inserting the occupation-number operator~\eqref{defNr} at level~$r$ 
between the two states introduces restrictions on the sum over partitions as follows,
\be \label{partsrestrict}
\bra{{\bf t}^+} N_r \ket{{\bf t}^-}  \= \sum_{r \,\in \, \mid \, \bmu \rangle} 
S_{\bmu}({\bf t}^+) \,  S_{\bmu}({\bf t}^-) \quad \Leftrightarrow \quad 
\bra{{\bf t}^+} (1-N_r) \ket{{\bf t}^-}  \= \sum_{r \,\notin \, \mid \, \bmu \rangle} 
S_{\bmu}({\bf t}^+) \,  S_{\bmu}({\bf t}^-) \,.
\ee
The restriction~$\ell(\mu)\le N$ in the sum~\eqref{ZNtlSS} translates to~$b_s < N$ 
in the Frobenius coordinates, which is to say that we sum over all partitions that 
do not contain any excitations around the vacuum of depth greater than~$N$ inside 
the Fermi sea (Figure~\ref{FiniteN}). 
From~\eqref{partsrestrict}, we have  
\be \label{ZNfreefer}
\wt Z_N({\bf t}^+, {\bf t}^-) \= \sum_{\ell(\bmu) \le N} S_{\bmu}({\bf t}^+) \,  S_{\bmu}({\bf t}^-) 
\= \bra{{\bf t}^+} \prod_{r < -N} (1-N_r) \ket{{\bf t}^-}  \,.
\ee
Upon expanding the infinite product, and using the expression for the number operator~\eqref{defNr},
we obtain
\be \label{ZNfreefersum}
\wt Z_N({\bf t}^+, {\bf t}^-) 
\= \sum_{m=0}^\infty (-1)^m  \sum_{N<r_1 < \dots < r_m}  \! \! \bra{{\bf t}^+}
\psi_{r_{1}} \psibar_{-r_1} \dots  \psi_{r_{m}} \psibar_{-r_m}  \ket{{\bf t}^-}  \,.
\ee

It now remains to calculate the~$2m$-point correlation functions in the 
expansion~\eqref{ZNfreefersum}. Let's first recall a much simpler problem, namely the 
evaluation of the same expectation values in the Dirac vacuum state. This is an elementary 
exercise involving moving all the positive energy modes to the right using the commutation 
relations~\eqref{psirsanticomm} as summarized by Wick's theorem, and it leads to 
the~$m \times m$ determinant of~2-point functions, which can be evaluated in free field theory. 
Thus, the only non-trivial feature here is the dependence on the states~$\ket{{\bf t}^\pm}$.
As it turns out, this can be reduced to a free field calculation using another piece of the 
formalism---also very familiar to physicists---namely the bosonization of the two-dimensional 
complex chiral fermion. 
We review and summarize the main steps in Appendix~\ref{App:bosonization}, and present 
the final answer for~$\wt Z_N$ which is written as sum over determinants. 

\begin{thm}\emph{\cite{BorOk} (Determinantal formula for~$\wt Z_N$)} \\
With the notations as above, we have 
\be \label{Zntlseries}
\frac{\wt Z_N({\bf t}^+, {\bf t}^-)}{\wt Z_\infty({\bf t}^+, {\bf t}^-)} \= 
\sum_{m=0}^\infty \, (-1)^m \, \! \!  \sum_{N<r_1 < \dots < r_m \atop r_i \in \IZ + \half}  
\det \bigl(\wt K(r_i, r_j \,; \, {\bf t}^+, {\bf t}^-) \bigr)_{i,j=1}^m \,,
\ee
where~$K(r_i, r_j \,; \, {\bf t}^+, {\bf t}^-)$, $r_i \in \IZ + \tfrac12$ are given by the generating function
\be \label{defkappatl}
\wt \kappa (z,w; \, {\bf t}^+, {\bf t}^-) \= 
\sum_{r,s \, \in \,\IZ+\half} \wt K(r,s \,; \, {\bf t}^+, {\bf t}^-) \, z^r \, w^{-s} 
\= \frac{J(z; \,{\bf t}^+, {\bf t}^-)}{J(w;\,{\bf t}^+, {\bf t}^-)} \; 
\frac{\sqrt{zw}}{z-w} \,, \qquad |w| < |z|  \,,
\ee
where
\be \label{defJ}
J(z; \,{\bf t}^+, {\bf t}^-) \= 
\exp\biggl(\, \sum_{k=1}^\infty \, \frac{1}{k} \, t^+_k \, z^k - \sum_{k=1}^\infty  \, \frac{1}{k} \, t^-_k \, z^{-k} \biggr) \,.
\ee
\end{thm}

\vskip 0.4cm

The expressions~\eqref{defkappatl}, \eqref{defJ} make it clear 
that~$\wt K(r,s \,; \, {\bf t}^+, {\bf t}^-)$, which is the 2-point function in the fermionic theory, 
has minimal degree at least~$r+s$ in the variables~${\bf t}^+, {\bf t}^-$ and, therefore, 
the~$m^\text{th}$ terms in~\eqref{Zntlseries} have minimal degree at 
least~$2mN+m(m+1)$. Since~$\wt Z_\infty$ has only non-negative powers of~$t_k^\pm$,~$k>0$, 
the previous statement about the minimal degree of the~$m^\text{th}$ term also holds 
for~$\wt Z_N({\bf t}^+, {\bf t}^-)$. Consequently, the coefficient of any given monomial 
in the variables~$(t^\pm_1,t^\pm_2, \dots)$ in the left-hand side receives contributions 
only from a finite number of terms on the right-hand side and therefore is well-defined.

\section{The giant graviton expansion  \label{sec:giants}}

In this section we return to the original problem posed by the integral~$Z_N({\bf g})$ 
defined in~\eqref{Uact} and the specialization~\eqref{indtrace}. 
Let us first take stock. We saw in Section~\ref{sec:HStrans} that~$Z_N({\bf g})$ is 
related to the auxiliary integral~$\wtZ_N({\bf t}^+, {\bf t}^- )$ defined in~\eqref{Uactaux} as 
\be \label{ZZtlrel}
Z_N({\bf g}) \= \vev{\wt Z_N}_{\bf g} \,,
\ee
where we recall the averaging operator~$\vev{ \cdot }_{\bf g}$ acts on power 
series~$f({\bf t}^+, {\bf t}^-)$ as  
\be \label{HStrans1}
\vev{ f }_{\bf g} \= \prod_{k=1}^\infty \int  \frac{\dd t_k^+ \, \dd t_k^-}{2 \pi k g_k} \, 
\rme^{-t_k^+\, t_k^-/k g_k} \, f({\bf t}^+, {\bf t}^-) \,.
\ee
At~$N=\infty$, the two integrals reduce to 
\be
Z_\infty({\bf g}) \=  \prod_{k=1}^\infty \frac{1}{1-g_k} \,, \qquad 
\wt Z_\infty({\bf t}^+, {\bf t}^-) \= \exp \biggl( \, \sum_{k=1}^\infty  \frac{1}{k} \, t^+_k \, t^-_k  \biggr) \,,
\ee
for which the relation~\eqref{ZZtlrel} can be easily verified. 

For finite~$N$ we have the formula~\eqref{Zntlseries} for the auxiliary matrix integral,
in terms of the 2-point function~$\wt K(r,s \,; \, {\bf t}^+, {\bf t}^-)$ given by~\eqref{defkappatl}.
Now we need to calculate the average~\eqref{ZZtlrel}. 
Note that, if we have~$t_k^\pm$ to be complex conjugate variables, and if the 
states~$\ket{{\bf t}^{-}}$ span a complete basis of states, then the required average 
is the expectation value 
\be
Z_N({\bf g}) \= \Tr \, \Bigl( \, \textstyle{\prod_{r < -N} (1-N_r)} \, \rho_{\bf g} \Bigr)
\ee
of the operator in the right-hand side of Equation~\eqref{ZNfreefer} in the quantum statistical state 
defined by the density operator 
\be
\rho_{\bf g} \= \exp(- {\bf t^+}{\bf t^-}/{\bf g}) \ket{{\bf t^-}} \bra{\bf t^+} \,,
\ee
with the trace and the exponential defined by the conventions in~\eqref{HStrans1}.

It is convenient to define a normalized transformation acting on each term 
in~\eqref{Zntlseries} as follows,
\be \label{defKg}
K^{(m)}_{\bf g}(r_1, \dots , r_m) \defeq
\frac{1}{Z_\infty({\bf g})} 
\vev{\wt Z_\infty \, \det \bigl(\wt K(r_i, r_j ) \bigr)_{i,j=1}^m }_{\bf g} \,.
\ee
We reach the formula for~$Z_N({\bf g})$ by summing~\eqref{defKg} over all~$m$-tuples 
of points in~$\IZ+\half$ as in~\eqref{Zntlseries}, and then summing over~$m=1,2,\dots$.
We have already seen that the~$m^\text{th}$ term in the expansion 
for~$\wt Z_N({\bf t}^+, {\bf t}^-)$ has minimal degree in~$({\bf t}^+, {\bf t}^-)$ 
at least~$2mN+m(m+1)$. 
From the action~\eqref{ttbargtrans} of the~H-S transformation on monomials, 
and the fact that~$Z_\infty({\bf g})$ has only non-negative powers of~$g_k$, we see 
that the~$m^\text{th}$ term in the expansion for~$Z_N({\bf t}^+, {\bf t}^-)$ 
has minimal degree in~${\bf g}$ at least~$mN+m(m+1)/2$, 
and therefore the sum over~$m$ 
is convergent in the power-series sense.  We have proved the following proposition.

\begin{proposition} \label{prop:ZNexp}
\emph{(A systematic expansion for~$Z_N$)}\\
The unitary matrix integral
\be 
Z_N({\bf g}) \= 
\int_{U(N)} \, \dd U\, \exp \biggl( \; \sum_{k=1}^\infty \, \frac{1}{k} \, g_k \,
 \Tr \, U^k \, \Tr \, U^{-k} \, \biggr) \,,
\ee
is given by the expansion 
\be \label{ZNexp}
\frac{Z_N({\bf g})}{Z_\infty({\bf g})} 
 \=  \sum_{m=0}^\infty \, G^{(m)}_N ({\bf g}) \,,
\ee
with~$G^{(0)}_N ({\bf g}) =1$ and 
\be \label{GmNsum}
G^{(m)}_N ({\bf g})  \=   (-1)^m 
\sum_{N<r_1 < \dots < r_m \atop r_i \in \IZ + \half} K^{(m)}_{\bf g}(r_1, \dots , r_m) \,, 
\qquad m \= 1,2,\dots \,,
\ee
where the functions~$K^{(m)}_{\bf g}$ are given in Equation~\eqref{defKg}. 
The minimal degree of the series~$G^{(m)}_N ({\bf g})$ is not less than~$mN+m(m+1)/2$. 
\end{proposition}

We call the~$G^{(m)}_N$ the~\emph{contribution from~$m$ giant gravitons} for reasons 
explained in the introduction.  
The above proposition gives a well-defined expansion for~$Z_N({\bf g})$ with each term
being the Hubbard-Stratonovich transformation of a determinant of two-point functions in our 
auxiliary theory. 
Performing the integrals involved in the transformation leads to further simplification. 
We now demonstrate this for the first non-trivial term in the expansion~\eqref{ZNexp}, 
namely the contribution of one giant graviton.

The contribution of one giant graviton is given in terms of 
the H-S transformation of the two-point function~$\wt K(r, s; \, {\bf t}^+, {\bf t}^-)$,
\be \label{Ktlrstrans}
\vev{\wt Z_\infty \, \wt K(r, s)}_{\bf g} \, \quad \text{with~$r=s$}.
\ee 
We consider the generating function of~\eqref{Ktlrstrans}, with one variable which tracks the 
power of~$r(=s)$. Recalling from~\eqref{ttbargtrans} that the H-S transform 
vanishes when the power of~${\bf t}^+$ and~${\bf t}^-$ are not equal, 
we see that one can sum over all values of~$(r,s)$ with two corresponding variables,
and the resulting function should only depend on one combination of the two variables. 
With this in mind, we consider 
\be \label{Ktlzwtrans1}
\begin{split}
& \sum_{r,s \in \IZ + \half}  z^r \, w^{-s} \, \vev{\wt Z_\infty \, \wt K(r, s)}_{\bf g} \
 \= \vev{\wt Z_\infty \;  \wt \kappa (z, w)}_{\bf g} \\
& \qquad \= \frac{\sqrt{zw}}{z-w}  \, \prod_{k=1}^\infty \int  \frac{\dd t_k^+ \, \dd t_k^-}{2 \pi k g_k} \,
\exp \biggl(\frac{1}{k} \Bigl( -\frac{t_k^+ \, t_k^-}{g_k} + t_k^+ \, t_k^-   + t_k^+(z^k - w^k) 
- t_k^-(z^{-k} - w^{-k}) \Bigr) \biggr) \,,
\end{split}
\ee
where we have interchanged the order of sum and averaging operations, and used~\eqref{defkappatl}.
Using the Gaussian integral formula~\eqref{Gaussianavg} we obtain, for~$ |w/z| <1$,
\be \label{Ktlzwtrans}
\begin{split}
& \sum_{r,s \in \IZ + \half}  z^r \, w^{-s} \, \vev{\wt Z_\infty \, \wt K(r, s)}_{\bf g} \\
& \qquad\qquad \= Z_\infty ({\bf g}) \, \frac{\sqrt{w/z}}{1-w/z} \,
\exp \biggl(\,\sum_{k=1}^\infty 
\frac{1}{k} \,\g_k \, \Bigl(-2 + \Bigl(\frac{w}{z}\Bigr)^k + \Bigl(\frac{w}{z}\Bigr)^{-k}  \Bigr)  \biggr) \,,
\qquad\qquad  
\end{split}
\ee 
where we have defined a new set of coupling constants,   
\be
\g_k \defeq \frac{g_k}{1-g_k} \,.
\ee
As anticipated above, the final expression for the transformed generating function~\eqref{Ktlzwtrans} is 
diagonal.
After dividing through by~$Z_\infty ({\bf g})$, we obtain the generating function for~$K^{(1)}_{\bf g}(r)$,
\be \label{K1rgen}
\sum_{r \in \IZ + \half}  \z^{-r}  \, K^{(1)}_{\bf g}(r) 
 \= \frac{\sqrt{\z}}{1-\z} \, 
\exp \biggl(\,\sum_{k=1}^\infty \frac{1}{k} \,\g_k \, \bigl(-2 + \z^k + \z^{-k}  \bigr)  \biggr) \,.
\ee 
We now see that the sum that we need in~\eqref{GmNsum}, 
\be \label{1giant}
G^{(1)}_N({\bf g}) \= - \sum_{N<r} K^{(1)}_{\bf g}(r) \,,
\ee
is equal to the negative of the~$\z^{-N}$ coefficient of the left-hand side of~\eqref{K1rgen} 
multiplied by~$\sqrt{\z}/(1-\z)$. 
Multiplying the right-hand side with the same factor thus gives us the 
corresponding generating function.
We have thus proved the following proposition. 

\begin{proposition} \label{prop:onegiant}
\emph{(Contribution of one giant)} \\
The generating series  
\be
\wh K(\z; \boldsymbol{\g}) \= - \sum_{N \in \IZ} \, \z^{-N} \, G^{(1)}_N({\bf g}) \,,
\ee
for the contribution~$G^{(1)}_N ({\bf g})$ of one giant-graviton to the expansion~\eqref{ZNexp}, equals 
\be
\wh K(\z; \boldsymbol{\g}) \= 
\frac{\z}{(1-\z)^2} \, \exp \biggl(\,\sum_{k=1}^\infty \frac{1}{k} \,\g_k \, \bigl(-2 + \z^k + \z^{-k}  \bigr)  \biggr) \,,
\ee
where~$\boldsymbol{\g} = (\g_1, \g_2, \dots)$ is given by~$\g_k = g_k/(1-g_k)$, $k=1,2,\dots$.
\end{proposition}

\vskip 0.4cm

\ndt {\bf Specialization to~$g_k=f(q^k)$}

\vskip 0.1cm

The index~\eqref{indtrace} is now easily evaluated by specializing the above considerations 
to couplings given in terms of a power series~$g_k = f(q^k)$. Proposition~\ref{prop:ZNexp} 
leads to a convergent expansion for a power series in~$q$, and Proposition~\ref{prop:onegiant} 
naturally leads to a new set of couplings and power series, 
\be
\g_k \= \wh f(q^k) \,, \qquad \wh f(q) \= \frac{f(q)}{1-f(q)} \,,
\ee
and the generating function
\be
\wh K_f(\z,q) \= \wh K \bigl(\z; (\wh f(q), \wh f(q^2), \dots) \bigr) \,.
\ee
We summarize our findings in terms of the following theorems.

\begin{thm} \label{thm:giants}
\emph{(Giant graviton expansion for the index)} \\
For any power series~$f(q)= \sum_{k =1}^\infty a_k \, q^k$ of order~$\a$, the matrix integral 
\be 
I^f_N(q) \= 
\int_{U(N)} \, \dd U\, \exp \biggl( \; \sum_{k=1}^\infty \, \frac{1}{k} \, f(q^k) \,
 \Tr \, U^k \, \Tr \, U^{-k} \, \biggr) 
\ee
admits a~$q$-expansion 
\be \label{giantgravexp}
\frac{I^f_N(q)}{I^f_\infty(q)} \= \sum_{m=0}^\infty \, G^{(m)}_{f,N}(q) \,, 
\qquad \II^f_\infty(q) =   \prod_{k=1}^\infty \frac{1}{1-f(q^k)} \,,
\ee
where $G^{(m)}_{f,N}(q)$, $m=0,1,2,\dots$ is given by the formula~\eqref{GmNsum} 
with~${\bf g} = (f(q), f(q^2), \dots)$, and has order at least~$\a m N+\a m(m+1)/2$ 
as a power series in~$q$.
\end{thm}

\begin{thm} \label{thm:onegiant}
\emph{(The contribution of one giant)} \\
With the set up of Theorem~\ref{thm:giants}, we have that 
$G^{(1)}_{f,N}(q)$ is the coefficient of~$\z^{-N}$ in the expansion of
\be \label{defKfzeta}
\wh K_f(\z,q) \= \frac{1}{(1-\z)\,(1-1/\z)} \, \prod_{n=1}^\infty \, 
\biggl( \frac{(1-q^n)^2}{(1-q^n \, \z) \,  (1-q^n / \z)} \biggr)^{\wh a_n} \,,
\ee
in the range~$|q| < |\z| < 1$, 
where the integers~$\wh a_n$ are summarized by their 
generating function~$\wh f(q) = \sum_{n=1}^\infty \, \wh a_n \, q^n$ 
given by
\be
\wh f(q) \= \frac{f(q)}{1-f(q)} \,.
\ee
\end{thm}

\vskip 0.4cm

\ndt {\bf Comments and implications\footnote{Further aspects of the above results, 
including some of the points below, are being studied in more detail in work in progress 
by the author with P.~Benetti-Genolini, G.~Eleftheriou, and S.~Garoufalidis.}}
\begin{enumerate}
\item The expansion of the infinite product~\eqref{defKfzeta} should be performed first in~$q$,  
as consistent with the range. The coefficient of~$q^m$ is seen to be a polynomial in~$\z, 1/\z$
of order~$m$. We can then reinterpret the expansion as a Laurent series in~$\z$, where the 
coefficient of~$\z^n$ is a~$q$-series (which has order at least~$|n|$). 
\item The bound given on the order of~$G^{(m)}_{f,N} (q)$ is not strict. For~$m=1$
the order is indeed~$\a(N+1)$, but  
the bound can be improved in general. 
Recall that the bound was derived by calculating the smallest power in~${\bf t}^+, {\bf t}^-$ 
in the elements of the matrix~$\bigl(\wt K(r_i, r_j \,; \, {\bf t}^+, {\bf t}^-) \bigr)_{i,j=1}^m$ 
and then translating the consequent bound on the determinant 
to powers of~$q$ after the integral transform. The point is that there can be cancellations
in the determinants for~$m \ge 2$. 
Indeed, the lowest term in the above process is the determinant of the 
matrix~$\bigl( q^{\a \,N+\a(i+j)/2} \bigr)_{i,j=1}^m$, which vanishes. The 
order of~$G^{(m)}_{f,N} (q)$ is, therefore, at least~$\a(mN+m(m+1)/2+1)$ for~$m \ge 2$. 
\item It is clear from the above considerations that the contributions of multiple giants will be 
Taylor coefficients of certain infinite products whose multiplicities are determined by~$f(q)$, 
in a manner similar to Theorem~\ref{thm:onegiant}.
\item The formula~\eqref{giantgravexp} is an exact formula which means that the full index is 
captured by bound states of giant-gravitons and gravitons. This is quite surprising from the 
AdS/CFT viewpoint, and is reminiscent of similar phenomena in~AdS$_3$ space~\cite{Eberhardt:2020bgq}. 
\item 
The gauge theory index~$I^f_N(q)$ at finite~$N$ can be interpreted holographically as follows. 
Say~$f(q) = a_1 q + a_2 q^2 + \dots$,~$a_1 \neq 0$. Then the first~$N$ coefficients are given 
by the multi-graviton formula~$I^f_\infty$, the next~$N$ coefficients are given by the contribution 
of one giant graviton (with multi-graviton fluctuations on top), and so on. 
\item The formula~\eqref{defKfzeta} is similar to the formula for the superconformal index with 
the power series~$\wh f(q)$ playing the role of the single-particle index. 
It should be interpreted as the single-particle index of the giant gravitons,
following the ideas in~\cite{Arai:2019xmp,Imamura:2021ytr, Gaiotto:2021xce}. 
The map $f \mapsto \wh f$ should be accounted for by the curvature of the D-branes in AdS space,
but it would be nice to understand this in detail. 
\item The structure of the expansion in Theorems~\ref{thm:giants},~\ref{thm:onegiant} 
points to a duality which interchanges~$m \leftrightarrow N$. This duality should be regarded as a   
manifestation of the open-closed-open duality of~\cite{GopOCO} in the BPS sector of the theory.
\item The final formula~\eqref{giantgravexp} expresses a microscopic quantity---the gauge theory 
partition function---as an ensemble average in a free fermionic theory or, equivalently,
an average over matrix model theories. 
It may serve as a good example to test ideas in the physics literature relating 
such averages to black holes~\cite{Maldacena:2016hyu}.
\end{enumerate}

\section{Examples}

\vskip 0.1cm

The statements in the theorems~\ref{thm:giants},~\ref{thm:onegiant} hold for an arbitrary power 
series~$f(q)$ of the type considered therein, and it is illuminating to check it numerically for a 
given power series of interest. In this section we illustrate the consequences of the theorems
for the BPS indices of the~$\CN=4$ super Yang-Mills theory given in~\eqref{1overnBPS}.

We use the following notation for the Pochhamer symbols, 
\be
(q)_n \= \prod_{k=1}^n \, (1-q^k)\,, \quad (q)_\infty \= \prod_{k=1}^\infty \, (1-q^k) \,,\quad 
(x;q)_\infty \= \prod_{k=0}^\infty (1-x q^k) \,.
\ee

\vskip 0.4cm

\ndt {\bf $\half$-BPS index}

\vskip 0.1cm

\ndt Here we specialize to~$f(q)=f_{1/2}(q)=q$. 
It is well-known that~$I^{1/2}_N(q)=1/(q)_N$, and this was also presented as a motivating 
example in~\cite{Gaiotto:2021xce}. We quickly work through the formula from the perspective of 
random partitions that we take here. The formula~\eqref{INqpartsum} gives
\be
\II^{1/2}_N(q) \= \sum_{\bl}  \, q^{|\bl|} \, \frac{1}{z_{\bl}} \,\sum_{\ell(\bmu)\le N}\, \chi^{\bmu}(\bl)^2  \,.
\ee
We now perform the sum over~$\bl$ for a fixed weight~$|\bl|=n$ and a fixed length~$\ell(\bmu)=k$ to obtain 
\be
\sum_{|\bl|=n} \frac{1}{z_{\bl}} \,\sum_{\ell(\bmu)= k} \, \chi^{\bmu}(\bl)^2 \= P(n,k) \,,
\ee
where~$P(n,k)$ is the partition of~$n$ into~$k$ parts. 
This leads to  
\be
\II^{1/2}_N(q) \= \sum_{n=0}^\infty \sum_{k=0}^N\,  P(n,k) \, q^n \= \frac{1}{(q)_N} \,,
\ee
which is a well-known formula.

In this case the giant graviton expansion can be fully understood. Firstly, we have
\be
\II^{1/2}_\infty(q) \= \frac{1}{(q)_\infty} \,,
\ee
which is indeed the multi-graviton index. Then we have
\be
\frac{\II^{1/2}_N(q)}{\II^{1/2}_\infty(q)} \=  \frac{(q)_\infty}{(q)_N} \= (q^{N+1};q)_\infty \,.
\ee
We use the identity (see e.g.~\cite{ZagDilog} Chapter 2, Proposition 2) 
\be
(x;q)_\infty \= \sum_{m=0}^\infty \, \frac{(-1)^m \, q^{m \choose 2}}{(q)_m} \, x^m 
\ee 
to obtain the complete giant-graviton expansion 
\be \label{Iqhalfexp}
\frac{\II^{1/2}_N(q)}{\II^{1/2}_\infty(q)}
\= \sum_{m=0}^\infty \, \frac{(-1)^m \, q^{m+1\choose 2}}{(q)_m} \, q^{mN} \,.
\ee

Now let us check our formula. We have 
\be
\wh f_{1/2}(q)\=\frac{q}{1-q} \= \sum_{n=1}^\infty \, q^n\,, \qquad 
\wh K_{\frac12}(\z,q) \= \frac{(q;q)_\infty^2}{(\z;q)_\infty \, (1/\z;q)_\infty} \,.
\ee
We expand~$\wh K$ in powers of~$q$ and then~$\z$, as explained above, to obtain the 
giant-graviton contributions
\be
\wh K_{\frac12}(\z,q) \= \sum_{N \in \IZ} \, \z^{-N} \, \wh G^{(1)}_{\frac12,N}(q)\,, \qquad |q|< |\z| <1 \,.
\ee
We find:\\
$N=1$
\be
\begin{split}
G^{(1)}_{\frac12,1}(q) & \=  -q^2 - q^3 - q^4 \,\textcolor{blue}{- \; q^6 - q^8 - q^9 - q^{10} - 2q^{12} - q^{15} + \dots} \,, \\
\frac{I^{1/2}_1(q)}{I^{1/2}_\infty(q)} -1 &\= -q^2 - q^3 - q^4 \,
\textcolor{blue}{+\; q^7 + q^8 + q^9 + q^{10} + q^{11} - q^{15} + \dots} \,. \\
\end{split}
\ee
$N=2$
\be
\begin{split}
G^{(1)}_{\frac12,2}(q) & \=  -q^3 - q^4 - q^5 - q^6  \,\textcolor{blue}{ - \; q^8 - q^{10} - 2q^{12} - q^{14}  + \dots} \,, \\
\frac{I^{1/2}_2(q)}{I^{1/2}_\infty(q)} -1 &\=  -q^3 - q^4 - q^5 - q^6  \,
\textcolor{blue}{+ \; q^9 + q^{10} + 2q^{11} + q^{12} + 2q^{13} + q^{14} +\dots} \,. 
\end{split}
\ee
$N=3$
\be
\begin{split}
G^{(1)}_{\frac12,3}(q) & \=  -q^4 - q^5 - q^6 - q^7 - q^8  \,\textcolor{blue}{  - \, q^{10} - q^{12} - q^{14} - q^{15} - q^{16} + \dots} \,, \\
\frac{I^{1/2}_3(q)}{I^{1/2}_\infty(q)} -1 &\=
-q^4 - q^5 - q^6 - q^7 - q^8  \,\textcolor{blue}{+\; q^{11} + q^{12} + 2q^{13} + 2q^{14} + 2q^{15} + 2q^{16} + \dots } \,.
\end{split}
\ee

We see that~$G^{(1)}_{\frac12,N}(q)$ has order~$N+1$, and 
agrees with the expansion~\eqref{Iqhalfexp} (recall that~$G^{(0)}_{\frac12,N}=1$),
up to~$\tO(q^{2N+4})$ as predicted.  

\vskip 0.4cm

\ndt {\bf $\frac18$-BPS index}

\vskip 0.1cm

\ndt Here we have~$f(q) = 2q/(1+q)$, for which the superconformal index is called 
the~\emph{Schur index}~\cite{Gadde:2011uv}. This index has been studied in quite some 
depth, including from the point of view of modular
transformations~\cite{Razamat:2012uv,Beem:2021zvt,Pan:2021mrw}. 
The following explicit expression for the Schur index was obtained in~\cite{Bourdier:2015wda}, 
\be  \label{1/8prefac}
\II^{1/8}_N(q)\= \frac{1}{(q;q)_\infty \, (q;q^2)_\infty} \,
 \sum_{m=0}^\infty \, (-1)^m \, \biggl( \binom{N+m}{N} + {N+m-1 \choose N} \biggr) \, 
 q^{m N + m^2} \,.
\ee
Firstly, we note that the prefactor in~\eqref{1/8prefac} is indeed equal to
\be
\frac{1}{(q;q)_\infty \, (q;q^2)_\infty} \= \prod_{n=1}^\infty \, \frac{1+q^n}{1-q^n} \= I_\infty^{1/8}(q)\,,
\ee
where the last equality follows from the formula~\eqref{Iinfty} with~$f^{1/8}(q)=2q/(1+q)$. 
Thus we have
\be \label{18BPSgiant}
\frac{\II^{1/8}_N(q)}{\II^{1/8}_\infty(q)}\=  
 \sum_{m=0}^\infty \, (-1)^m \, \biggl( \binom{N+m}{N} + {N+m-1 \choose N} \biggr) \, 
 q^{m N + m^2} \,,
\ee
which is in the form of the giant graviton expansion. 

Now let us check our formula. We have 
\be
\wh f_{1/8}(q)\=\frac{2q}{1-q} \=  \= 2\sum_{n=1}^\infty \, q^n \,, \qquad 
\wh K_{\frac18}(\z,q) \= \frac{1}{(1-\z)(1-1/\z)} \frac{(q;q)_\infty^4}{(\z q;q)_\infty^2 \, (q/\z;q)_\infty^2} \,.
\ee
We find, upon expanding~$\wh K_{\frac18}$, the contributions of~$N=1,2,3$:
\be
\begin{split}
G^{(1)}_{\frac18,1}(q) & \= -3q^2 - 5q^6 - 7q^{12} - 9q^{20} + \tO(q^{21}) \,, \\
G^{(1)}_{\frac18,2}(q) & \= -4q^3 - 6q^8 - 8q^{15} + \tO(q^{21}) \,, \\
G^{(1)}_{\frac18,3}(q) & \= -5q^4 - 7q^{10} - 9q^{18} + \tO(q^{21}) \,.
\end{split}
\ee
We can check that~$G^{(1)}_N(q)$ has order~$N+1$ and 
agrees with the expansion~\eqref{18BPSgiant} up to~$\tO(q^{2N+4})$
as predicted.

\vskip 0.4cm

\ndt {\bf $\frac{1}{16}$-BPS index}

\vskip 0.1cm

\ndt Here we have
\be
\begin{split}
f_{1/16}(q) & \=   1-\frac{(1-q^2)^3}{(1-q^3)^2} \= \frac{q^2(3+4q+2q^2)}{(1+q+q^2)^2}  \,, \\
& \=  3 \sum_{n=1}^\infty \, n \, q^{3n-1} -2  \sum_{n=1}^\infty  \, q^{3n} - 3 \sum_{n=1}^\infty \, n \, q^{3n+1} \,,
\qquad \qquad \qquad  
\end{split}
\ee
\be
\begin{split}
\wh f_{1/16}(q) & \= \frac{(1-q^3)^2}{(1-q^2)^3}-1 \=\frac{q^2(3+4q+2q^2)}{(1-q)(1+q)^3}  \,, \\
&\= \sum_{n=1}^\infty \, (n^2+2) \, q^{2n} -\sum_{n=1}^\infty \, n(n-1) \, q^{2n-1} \,. 
\qquad \qquad \qquad\qquad  
\end{split}
\ee
In this case  no explicit formula (e.g.~of the type~\eqref{1/8prefac}) is known which does not involve
the evaluation of multi-dimensional contour integrals~\eqref{Uact} or the evaluation of symmetric group 
characters~\eqref{INqpartsum}. Using the latter method the index up to~$\tO(q^{70})$ was  
calculated in~\cite{Murthy:2020rbd}. Here~$f_{1/16}(q) = \tO(q^2)$, 
so that~$\a=2$ in the notation of Theorem~\ref{thm:giants}. This means 
that~$G^{(1)}_{{\frac{1}{16},N}}(q) = \tO(q^{2(N+1)})$, and so the non-trivial check with the first 70 terms 
can be made for up to~$N=34$. 
We have verified the statement of our theorem that~$G^{(1)}_{{\frac{1}{16},N}}(q)$ 
agrees with the expansion~\eqref{18BPSgiant} up to~$\tO(q^{4N+8})$ up to this rank. We present the first 
three cases below. We note that the functions~$\wh K_{\frac{1}{16}}(\z,q)$ are infinite products 
of terms of the type~$(1-q^n)^{n}$ and~$(1-q^n)^{n^2}$, generalizing the MacMahon function, 
such functions are being studied in work in progress by the author with S.~Garoufalidis and D.~Zagier. 

\vskip 0.4cm

\ndt $N=1$
\be
\begin{split}
G^{(1)}_{\frac{1}{16},1}(q) & 
\= -6q^4 + 6q^5 - 3q^6 - 6q^7 + 21q^8 - 36q^9 + 27q^{10} + 30q^{11}  \textcolor{blue}{-148q^{12} + 270q^{13} } \\
& \qquad \textcolor{blue}{- 336q^{14} + 202 q^{15} + 348 q^{16} - 1392 q^{17}  + \dots } \,, \\
\frac{I^{1/16}_1(q)}{I^{1/16}_\infty(q)} -1 & 
\= -6q^4 + 6q^5 - 3q^6 - 6q^7 + 21q^8 - 36q^9 + 27q^{10} + 30q^{11}  \textcolor{blue}{-92q^{12} + 132q^{13} } \\
& \qquad \textcolor{blue}{- 90q^{14} - 106q^{15} + 369q^{16} - 444q^{17} + \dots } \,.  \\
\end{split}
\ee
$N=2$
\be
\begin{split}
G^{(1)}_{\frac{1}{16},2}(q)  & \= 
-10q^6 + 12q^7 - 9q^8 + 21q^{10} - 54q^{11} + 83q^{12} - 102q^{13} + 72q^{14} + 128q^{15}  \\
& \qquad \textcolor{blue}{- 585q^{16} + 1122q^{17} + \dots }\,,  \\
\frac{I^{1/16}_2(q)}{I^{1/16}_\infty(q)} -1 & \= 
-10q^6 + 12q^7 - 9q^8 + 21q^{10} - 54q^{11} + 83q^{12} - 102q^{13} + 72q^{14} + 128q^{15}  \\
& \qquad \textcolor{blue}{- 459q^{16} + 744q^{17}  + \dots } \,.
\end{split}
\ee
$N=3$
\be
\begin{split}
G^{(1)}_{\frac{1}{16},3}(q)  &\=  -15q^8 + 20q^9 - 18q^{10} + 12q^{11} + 10q^{12} - 54q^{13} 
+ 111q^{14} - 190q^{15} + 279q^{16} \\
& \qquad  - 288q^{17} + 49q^{18} + 630q^{19} \textcolor{blue}{ - 1905q^{20} + 3658q^{21} + \dots } \,,\\ 
\frac{I^{1/16}_2(q)}{I^{1/16}_\infty(q)} -1 & \= 
 -15q^8 + 20q^9 - 18q^{10} + 12q^{11} + 10q^{12} - 54q^{13} + 111q^{14} - 190q^{15} + 279q^{16} \\
& \qquad - 288q^{17} + 49q^{18} + 630q^{19} \textcolor{blue}{- 1653q^{20} + 2790q^{21} + \dots } \,.
\end{split}
\ee

\vskip 1cm

\section*{Acknowledgements}

I would like to thank Dionysios Anninos, Pietro Benetti-Genolini, Nadav Drukker, Giorgos Eleftheriou, 
Greg Moore, Stavros Garoufalidis, Rajesh Gopakumar, Sanjaye Ramgoolam, Hassaan Saleem, Gerard Watts, and Don Zagier  
for interesting and useful conversations about topics discussed in this paper and for comments on an earlier draft.  
This work is supported by the ERC Consolidator Grant N.~681908, ``Quantum black holes: A microscopic 
window into the microstructure of gravity'', and by the STFC grant ST/P000258/1. 

\vskip 1cm

\appendix

\section{The matrix integral as a sum over partitions \label{App:sumparts}}

We use the notations and conventions of~\cite{Macdonald}.
We denote partitions as~$\bl = (\lambda_1, \lambda_2,\dots)$ 
with~$\lambda_1 \ge \lambda_2 \ge \dots $, or in the frequency representation 
as~$\prod_{i \ge 1} i^{\, r_i} =1^{r_1} \, 2^{r_2} \, \dots$. The number of parts of a partition 
(or length) is the number of non-zero~$\lambda_i$ or, equivalently, $\ell(\bl) \= \sum_{i}  r_i$. 
The weight of the partition~$|\bl| = \sum_{j\ge1} \lambda_j \= \sum_{i\ge1} \, i \, r_i$. 
The first step is to expand the exponential in~\eqref{Uact}, which gives a sum of 
products of traces of powers of the unitary matrix and powers of its inverse. 
Each such product~$\prod_{i \ge 1} \, \bigl(\Tr \, U^i \bigr)^{r_j}$ is labelled by the  
partition~$\bl = \prod_{i\ge 1} i^{\, r_i}$ written in the frequency representation, and we 
denote such a product by~$\CO_{\bl}(U)$ (and similarly we have~$\CO_{\bl}(U^{-1})$.)
Next we write these powers in terms of~$U(N)$ group characters~\cite{FultonHarris}.
Recall that the representations of~$U(N)$ and those of the symmetric group~$S_N$ are both 
labelled by partitions of~$N$. We denote the corresponding characters as~$\wt\chi_{\bmu}$
and~$\chi^{\bmu}$, respectively. 
We then use the Frobenius formula for~$U(N)$,
\be \label{Frobfor}
\CO_{\bl} (U) \= \sum_{\ell(\bmu)\le N} \, \wt \chi_{\bmu}(U) \, \chi^{\bmu}(\bl) \,.
\ee
Here, and below, all sums over partitions run over all partitions
with any restrictions being indicated (as in the sum over~$\bmu$).

Using the first orthogonality relation of the group characters of~$U(N)$, i.e.,
\be
\int \, DU\,   \wt \chi_{\bmu}(U) \, \wt \chi_{\bmu'}(U^{-1})  \= \delta_{\bmu \bmu'} \,,
\ee
we obtain
\be \label{upupbarint}
\int \, DU\, \CO_{\bl} (U) \, \CO_{\bl} (U^{-1})  
\= \sum_{\ell(\bmu)\le N}  \, \chi^{\bmu}(\bl) \; \overline{\chi^{\bmu}(\bl)} 
\= \sum_{\ell(\bmu)\le N}  \, \chi^{\bmu}(\bl)^2 \,. 
\ee 
Here the second equality is a consequence of the fact that the characters of the symmetric group
are real (actually integers).
Upon putting these steps together, we obtain
\be \label{ZNpartsum1}
\begin{split}
Z_N({\bf g}) & \= \int \, DU\,  \sum_{\bl}\frac{ {\bf g}^{\bl}}{z_{\bl}} \; \CO_{\bl} (U) \, \CO_{\bl}(U^\dagger)  \\
& \= \sum_{\bl} \, \frac{ {\bf g}^{\bl}}{z_{\bl}} \,
 \sum_{\ell(\bmu)\le N}  \, \chi^{\bmu}(\bl)^2  \,,
 \end{split}
\ee
where, for any partition~$\bl$ as above, we use the notations
\be
{\bf g}^{\bl} \defeq  \prod_{j\ge 1} \, g_{\lambda_j} \= \prod_{i\ge 1} \, g_i^{r_i} \,, \qquad 
z_{\bl} \defeq \prod_{i} \, r_i ! \, i^{r_i} \,. 
\ee

\section{Review of bosonization and fermionic correlation functions \label{App:bosonization}}

We first recall the basic ideas and equations of bosonization that we need here~\cite{Ginsparg:1988ui}
(we use the conventions of~\cite{Polchinski:1998rr}).
We introduce a real chiral bosonic field~$X(z)$ which obeys the operator product expansion 
\be \label{bosonOPE} 
\p X(z) \, \p X(w) \; \sim \; -\frac{1}{(z-w)^2} + \tO(1) \,.
\ee
The field is expanded in oscillator modes as 
\be
 \p X (z) \= - \ii \,\sum_{n\in\IZ} \frac{\a_n}{z^{n+1}} \,. 
\ee
The oscillator modes obey~${\a}^\dagger_n = \a_{-n}$, and 
the OPE~\eqref{bosonOPE} is equivalent to the commutation relations
\be
\bigl[ \, \a_n \, , \, \a_m \, \bigr]  \= n \, \delta_{n+m,0} \,.
\ee
We call~$\a_n$ bosonic creation operators for~$n<0$ and annihilation operators for~$n>0$.
The bosonic vacuum is defined to be the state 
\be \label{defbosvac}
\a_n \ket{0} \= 0 \,, \qquad n > 0 \,.
\ee
States in the bosonic Fock space are spanned by an arbitrary number of creation operators acting
on the vacuum state.
The mode~$\a_0$ corresponds to the momentum in field space and needs to be 
quantized separately. As it turns out, we do not need to deal with this zero mode for our purposes,
and therefore we will only discuss the modes~$n$ in~$\IZ^* = \IZ \, \backslash \, \{ 0\}$ in the 
formulas below. We will also not discuss cocycles in any of the vertex operators,  
as we do not need them for our purposes.

The basic statement of bosonization is that the free complex chiral fermion discussed in 
Section~\ref{sec:determinants} is equivalent to the free chiral boson with the relations
\be \label{psiXrel}
\psi(z) \= : \rme^{+\ii \, X(z)} : \,, \qquad \psibar(z) \=  :\rme^{-\ii \, X(z)} : \,,
\ee
where the~$: \cdot :$ indicates normal ordering. Equivalently, we can present the relation between the 
fermionic and the bosonic modes, 
\be
\a_n \= \sum_{r \in \IZ + \half} \, \psibar_{n-r} \, \psi_r \,, \qquad n \in \IZ^*  \,.
\ee
The bosonic vacuum maps to the Dirac vacuum.
We will also need the commutation relations of the bosonic oscillator modes with the fermions,
which follow from~\eqref{psiXrel},
\be
\bigl[ \, \a_n \, , \, \psi(z) \, \bigr] \= z^n \, \psi(z) \,, \qquad 
\bigl[ \, \a_n \, , \, \psibar(z) \, \bigr]  \= - z^n \, \psibar(z) \,, 
\qquad n \in \IZ^* \,.
\ee

\vskip 0.4cm

Next we write the state~$\ket{\bf t}$ defined in~\eqref{defkett} in terms of the bosonic variables.
This needs some basics of symmetric function theory~\cite{Macdonald}. 
An important role in the following is played by the following vertex operators
\be
\G_{\pm} ({\bf t}) \defeq \exp\biggl(\,\sum_{k=1}^\infty \frac{1}{k} \, t_k \, \a_{\pm k} \biggr) \,, 
\qquad {\bf t} = (t_1,t_2,\dots) \,,
\ee
which are Hermitian conjugates of each other~$\G_+({\bf t})^\dagger = \G_-({\bf t})$. 
They obey the relations 
\be\label{Gpmcomm}
\G_+({\bf t})\, \G_+({\bf t}') \= \exp \biggl( \, \sum_{k=1}^\infty \frac{1}{k} \, t_k \, t'_k  \biggr) \, 
\G_+({\bf t}') \,  \G_+({\bf t}) \,,
\ee
and 
\be \label{Gpmpsicomm}
\begin{split}
\G_{\pm}({\bf t})\, \psi(z) & \=  \exp\biggl(\, \sum_{k=1}^\infty \frac{1}{k} \, t_k \, z^{\pm k} \biggr) \, 
\psi(z) \,  \G_{\pm}({\bf t}) \,, \\
\G_{\pm}({\bf t})\, \psibar(z) & \=  \exp\biggl(\,- \sum_{k=1}^\infty \frac{1}{k} \, t_k \, z^{\pm k} \biggr) \, 
\psibar(z) \,  \G_{\pm}({\bf t}) \,.
\end{split}
\ee

\vskip 0.2cm

Note that the factors appearing in the equations~\eqref{Gpmpsicomm} are related to the complete 
homogeneous symmetric functions as follows,
\be
 \exp\biggl(\, \sum_{k=1}^\infty \frac{1}{k}\,  t_k \, z^k \biggr) \= \sum_{n=0}^\infty \, z^n \, H_n({\bf t}) \,.
\ee
Here~$H_n({\bf t}) = h_n({\bf x})$ for~$t_k = \sum_i x_i^k$, $k = 1,2,\dots $, with 
$h_n$ being the complete homogeneous symmetric functions in the alphabet~${\bf x}$~\cite{Macdonald}. 
It then follows from the relations~\eqref{Gpmpsicomm}
that the state 
\be
\ket{\bmu} \= \prod_{i=1}^s \psi_{-a_i} \, \psibar_{-b_i} \ket{0} \,,
\ee
as defined in~\eqref{partstate}, obeys 
\be \label{muketSchur}
\bra{\bmu}\G_-({\bf t}) \ket{0} 
\= \det \bigl(H_{{\mu_i}-i+j}({\bf t}) \bigr) \= S_{\bmu}({\bf t}) \,.
\ee
The functions~$S_{\bmu}({\bf t})$ appearing on the right-hand side are the Schur functions
of~${\bf x}$ when we substitute~$t_k \= \sum_i x_i^k$, and the last equality in~\eqref{muketSchur}
is the Jacobi-Trudy identity. 
The functions~$S_{\bmu}({\bf t})$ are precisely the functions defined as
\be  
S_{\bmu}({\bf t}) \= \sum_{\bl} \, \frac{{\bf t}^{\bl}}{z_{\bl}} \, \chi^{\bmu}(\bl) \,,
\ee
in~\eqref{schurdef}, as discussed below that equation. 

Upon combining~\eqref{muketSchur} and the definition~\eqref{defkett}, we obtain  
the statement that the wavefunction of the state~$\G_-({\bf t}) \ket{0}$ in the partition basis 
is the Schur function, i.e., 
\be \label{tisSchur}
\ket{{\bf t}}  \= \sum_{\bmu} S_{\bmu}({\bf t}) \ket{{\bmu}} \= \G_-({\bf t}) \ket{0}   \,.
\ee

\vskip 0.4cm

Next we derive the expression~\eqref{Zntlseries}.   
Firstly, we notice that 
the~${\bf t}^\pm$-dependence of the state in~\eqref{ZNfreefersum} can be transferred to 
a~${\bf t}^\pm$-dependence of the operators. 
From the relations~\eqref{tisSchur},~\eqref{defbosvac},~\eqref{Gpmcomm},
and the identification~\eqref{ZNtlinfty}, we obtain 
\be \label{mptfer}
\frac{1}{\wt Z_\infty({\bf t}^+, {\bf t})} \, 
\bra{{\bf t}^+} \psi_{r_{1}} \psibar_{-r_1} \dots  \psi_{r_{m}} \psibar_{-r_m}  \ket{{\bf t}^-} \=
 \bra{0} \Psi_{r_{1}} \Psibar_{-r_1}  \dots   \Psi_{r_{m}}\Psibar_{-r_m} \ket{0} \,,
\ee
where we have defined the dressed fermions
\be \label{defPsir}
\begin{split}
& \Psi_r  \= \Psi_r ({\bf t}^+, {\bf t}) \= \GG  \, \psi_r \, \GG^{-1} \,, \qquad 
\Psibar_r  \= \Psibar_r({\bf t}^+, {\bf t}) \= \GG \, \psibar_r \, \GG^{-1} \,, \\
 & \qquad \qquad\qquad  \GG  \= \GG({\bf t}^+, {\bf t})  \= \G_+({\bf t}^+) \, \G_-({\bf t}^-)^{-1} \,.
\end{split}
\ee
It is easy to check that~$\Psi_r$ and $\Psibar_s$ anicommute when~$r \neq s$, and therefore 
the right-hand side of~\eqref{mptfer} can be evaluated as usual by Wick's theorem to give the 
determinant of 2-point functions. 
We have
\be
 \bra{0} \Psi_{r_{1}} \Psibar_{-r_1}  \dots   \Psi_{r_{m}}\Psibar_{-r_m} \ket{0} \= 
 \det \Bigl(  \bra{0} \Psi_{r_i} \Psibar_{-r_j} \ket{0}\Bigr)_{i,j=1}^m \,.
\ee

Thus the problem has reduced to evaluating the 2-point functions
\be
 \wt K(r,s;\,{\bf t}^+, {\bf t})  \defeq  \bra{0} \Psi_{r} ({\bf t}^+, {\bf t}^-) \, \Psibar_{-s} ({\bf t}^+, {\bf t}^-) \ket{0} \,,
\ee
which 
we collect in the generating function
\be
\begin{split}
\wt \kappa (z,w; \, {\bf t}^+, {\bf t}^-) & \defeq \sum_{r,s \, \in \IZ+\half} \wt K(r,s; \, {\bf t}^+, {\bf t}^-) \, z^{-r} \, w^{s} \,,\\
& \=  \sqrt{zw} \; \bra{0} \GG({\bf t}^+, {\bf t}^-) \, \psi(z) \, \psibar (w) \, \GG({\bf t}^+, {\bf t}^-)^{-1} \ket{0} \,,
\end{split}
\ee
where we have used the definitions~\eqref{defPsir} in going to the second line. 
From the quasi-commutation relations~\eqref{Gpmcomm},~\eqref{Gpmpsicomm}, we obtain
\be \label{kappa2pt}
\wt \kappa (z,w; \, {\bf t}^+, {\bf t}^-)  
 \= \sqrt{zw} \, \frac{J(z; \,{\bf t}^+, {\bf t}^-)}{J(w;\,{\bf t}^+, {\bf t}^-)} \; \bra{0} \psi(z) \, \psibar (w) \ket{0}  \,,
\ee
where
\be
J(z; \,{\bf t}^+, {\bf t}^-) \= \exp\biggl(\, \sum_{k=1}^\infty \, \frac{1}{k} \, t^+_k \, z^k - 
\sum_{k=1}^\infty  \, \frac{1}{k} \, t^-_k \, z^{-k} \biggr) \,.
\ee
The fermionic free-field correlator in~\eqref{kappa2pt} is standard and follows easily from the original 
algebra~\eqref{psiOPE} or~\eqref{psirsanticomm},
so that 
\be
\wt \kappa (z,w; \, {\bf t}^+, {\bf t}^-)  
\= \frac{J(z; \,{\bf t}^+, {\bf t}^-)}{J(w;\,{\bf t}^+, {\bf t}^-)} \; \frac{\sqrt{zw}}{z-w} \,, \quad |w| < |z|  \,.
\ee

Upon putting these formulas together, we obtain the formula~\cite{BorOk} for~$\wt Z_N$ 
as a sum over determinants, 
\be 
\frac{\wt Z_N({\bf t}^+, {\bf t}^-)}{\wt Z_\infty({\bf t}^+, {\bf t}^-)} \= 
\sum_{m=0}^\infty \, (-1)^m \, \! \!  \sum_{N<r_1 < \dots < r_m \atop r_i \in \IZ + \half}  
\det \bigl(\wt K(r_i, r_j; \, {\bf t}^+, {\bf t}^-) \bigr)_{i,j=1}^m \,.
\ee


\bibliographystyle{alpha}

\newcommand{\etalchar}[1]{$^{#1}$}

\end{document}